\documentclass[fleqn,usenatbib]{mnras}

\usepackage{newtxtext,newtxmath}

\usepackage[T1]{fontenc}

\DeclareRobustCommand{\VAN}[3]{#2}
\let\VANthebibliography\thebibliography
\def\thebibliography{\DeclareRobustCommand{\VAN}[3]{##3}\VANthebibliography}

\usepackage{graphicx}	
\usepackage{amsmath}	
\usepackage[dvipsnames]{xcolor} 
\usepackage{lscape} 
\usepackage{longtable} 
\usepackage{subfig}

\usepackage{color, colortbl} 
\definecolor{Gray}{gray}{0.9}

\title[X-ray AGN in Nearby Galaxies]{The Incidence of X-ray selected AGN in Nearby Galaxies}

\author[K. L. Birchall et al.]{
Keir L. Birchall$^1$,\thanks{E-mail: klb69@leicester.ac.uk}
M. G. Watson$^1$,
J. Aird$^{2,1}$,
and R. L. C. Starling$^1$
\\
$^1$School of Physics \& Astronomy, University of Leicester, University Road, Leicester LE1 7RH, UK\\
$^2$Institute for Astronomy, University of Edinburgh, Royal Observatory, Edinburgh EH9 3HJ, UK\\
}

\date{Accepted 2021 December 3. Received 2021 October 20; in original form 2021 January 29}

\pubyear{2020}

\begin{document}

\label{firstpage}
\pagerange{\pageref{firstpage}--\pageref{lastpage}}
\maketitle

\begin{abstract}
We present the identification and analysis of an unbiased sample of AGN that lie within the local galaxy population. Using the MPA-JHU catalogue (based on SDSS DR8) and 3XMM DR7 we define a parent sample of 25,949 local galaxies ($z \leq 0.33$). After confirming that there was strictly no AGN light contaminating stellar mass and star-formation rate calculations, we identified 917 galaxies with central, excess X-ray emission likely originating from an AGN. We analysed their optical emission lines using the BPT diagnostic and confirmed that such techniques are more effective at reliably identifying sources as AGN in higher mass galaxies: rising from 30\% agreement in the lowest mass bin to 93\% in the highest. We then calculated the growth rates of the black holes powering these AGN in terms of their specific accretion rates ($\propto L_X/M_*$). Our sample exhibits a wide range of accretion rates, with the majority accreting at rates $\leq 0.5\%$ of their Eddington luminosity. Finally, we used our sample to calculate the incidence of AGN as a function of stellar mass and redshift. After correcting for the varying sensitivity of 3XMM, we split the galaxy sample by stellar mass and redshift and investigated the AGN fraction as a function of X-ray luminosity and specific black hole accretion rate. From this we found the fraction of galaxies hosting AGN above a fixed specific accretion rate limit of $10^{-3.5}$ is constant (at $\approx 1\%$) over stellar masses of $8 < \log \mathrm{M_*/M_\odot} < 12$ and increases (from $\approx 1\%$ to $10\%$) with redshift.  

\end{abstract}


\section{Introduction}
The degree to which a supermassive black hole (SMBH) and its host galaxy interact and affect each other's evolution is an important and contested question. Among the well-studied and constrained examples of co-evolution are the relations between an SMBH's mass and the velocity dispersion, mass and luminosity of the host galaxy's classical bulge. \citep{Magorrian98, FerrareseMerritt00, Gebhardt00}.  Such relationships imply that the SMBH and bulge co-evolve by regulating each others growth \citep{KormendyHo13}. SMBH growth can also be identified through a range of electromagnetic signatures released when they undergo periods of accretion. Tracing the features of active galactic nuclei (AGN) activity across cosmic time has highlighted more evidence of co-evolution. For example, total AGN accretion rate density is highly correlated with changes in star-formation rate (SFR) density \citep{MadauDickinson14, Delvecchio14, Aird15}; the effect of AGN on their host galaxies in the form of feedback is associated with the quenching of star formation \citep{DiMatteo05, Fabian12,Greene20}; and feedback could also explain the observed bi-modality in colour-magnitude and colour-mass space \citep{Baldry04, Martin07, Schawinski14}.\\

It is still unclear what mechanism is the main driver behind getting gas and dust down into the centre \citep{AlexanderHickox12}. As discussed in a number of studies, taking large samples of galaxies is a useful way of constraining the AGN fuelling mechanism as it allows us to smooth out stochastic differences and try to identify trends \citep{Aird13, Hickox14, Delvecchio20}. 
The most complete AGN samples are typically found by using X-ray surveys as they can identify AGN across large redshift ranges and down to relatively low luminosities where non-AGN emission may dominate \citep{BrandtAlexander15}. Studies that have adopted this approach suggest that AGN are much more likely to be found in higher mass galaxies, and in galaxies at higher redshifts \citep{Xue10, Haggard10, Mendez13, HernanCaballero14,WilliamsRottgering15}.\\
	
Whilst these surveys employ large samples of AGN, they are subject to several observational biases. Firstly, most X-ray detection techniques define AGN as objects that exceed absolute luminosity thresholds, or more generally dominate galaxy emission \citep{BrandtAlexander15}, causing them to miss the lower luminosity, less efficiently accreting black holes typically found in dwarf galaxies. Any low mass black holes that are detected are the much rarer, more actively accreting ones. However, over the past decade there has been a huge increase in the number of AGN detected in dwarf galaxies aided, in part, by adapting these techniques. X-ray selected AGN in dwarf galaxies are identified through isolating a centrally located excess emission that cannot be explained by other sources, including X-ray binaries and hot gas \citep{Lemons15, Paggi16, Pardo16, Baldassare17,Mezcua18, Birchall20}. Such a change in detection method allows us to overcome the bias towards actively accreting, higher mass black holes associated with the absolute X-ray luminosity thresholds and helps challenge the assumption that AGN are less likely in lower mass galaxies.\\

Secondly, a lot of these studies make little attempt to understand and correct for the limitations of the survey data used. \cite{Aird12} was among the first AGN studies to attempt to overcome these observational limitations and apply completeness corrections to their sample. By applying these corrections to a sample of galaxies out to $z \approx 1$, from the Prism Multi-object Survey \citep{Coil11,Cool13}, they were able to determine the probability of finding an AGN as a function of various host galaxy properties. \cite{Aird12} showed that the probability of hosting an AGN can be described as a power law of X-ray luminosity and specific black hole accretion rate (sBHAR). Both X-ray luminosity and sBHAR distributions were found to be consistent over a wide range of stellar masses but with normalisations that dropped rapidly with decreasing redshift. Similar approaches to \cite{Aird12} have been adopted and adapted by subsequent AGN population studies \citep{Bongiorno12, Georgakakis17, Aird17, Aird18, Birchall20}.\\

By combining the large galaxy sample available in the SDSS and the X-ray data from XMM-Newton Serendipitous Sky Survey, \cite{Birchall20} found power law distributions describing the AGN incidence in dwarf galaxies as a function of X-ray luminosity and sBHAR. From this we identified a redshift-independent fraction out to $z \sim 0.7$. We also found a weak increase in AGN fraction with stellar mass, however it was also consistent with being flat for the range probed. In this paper, we take the methods used in \cite{Birchall20}, remove the mass threshold and investigate the X-ray luminosity and sBHAR distributions for this expanded sample of local AGN. \cite{Georgakakis11} have previously calculated the space density of X-ray AGN with $\mathrm{\log_{10}L_X}>41$ at $0.03<z<0.2$ based on a sample constructed using both XMM and SDSS observations \citep{GeorgakakisNandra11}. However, there have been thousands of new XMM observations since this work was produced. Thus, with this expanded survey and our new methods to account for sample incompleteness we aim to identify whether there is any connection between AGN incidence and stellar mass or redshift. \\

The paper is structured as follows. First, we describe the construction of our AGN sample (§\ref{sec:data}), making sure to verify that large non-stellar nuclear emission does not affect our host galaxy measurements. Then, we assess the extent to which our X-ray selected sample overlaps with BPT selection (§\ref{sec:BPT}). Next, we investigate the rate at which the black holes powering our AGN are accreting material (§\ref{sec:sBHAR}). With this information we then create our X-ray luminosity and sBHAR distributions (§\ref{sec:Prob_Dists}) and use them to calculate how the AGN fraction varies with stellar mass and redshift (§\ref{sec:fractions}). Throughout, we assume Friedmann-Robertson-Walker cosmology: $\Omega = 0.3$, $\Lambda = 0.7$ and $H_0 = 70\mathrm{\ km\ s^{- 1} Mpc}^{- 1}$. \\

\section{Data \& Sample Selection}
\label{sec:data}
As this paper aims to build upon the work started in \cite{Birchall20}, we continue to use the catalogues described therein. In this section, we will provide a brief summary of these catalogues - please refer to the original paper for a more detailed description. In addition, we will outline the new, more rigorous position-matching process needed to efficiently match our expanded sample. \\

Optical photometry and spectroscopy covering about a quarter of the sky ($ \approx 9274\ \mathrm{deg^2}$) can be found in the Sloan Digital Sky Survey Data Release 8 (SDSS DR8). Any SDSS DR8 spectra classified by the pipeline as a galaxy was then considered for further processing by the MPA-JHU value-added catalogue \footnote{Available at \href{http://www.mpa-garching.mpg.de/SDSS/DR7/}{http://www.mpa-garching.mpg.de/SDSS/DR7/}}. It provides estimates of galaxy properties such as stellar mass, star formation rate (SFR) and emission line fluxes, for 1,472,583 objects in this release. From this catalogue we isolate the 835,861 objects with a "good" photometric reliability flag. This indicates valid results from MPA-JHU's photometric fits which indicates useable mass and star-formation rate (SFR) measurements.\\

Stellar masses are calculated using the ugriz photometry from the full extent of the galaxy, and UV data from GALEX \citep{Martin05}. The total stellar mass is calculated by fitting to model magnitudes based on the method described in \cite{Kauffmann03}. MPA-JHU stellar masses were found to be largely consistent with those from the Galex-SDSS-WISE Legacy Catalogue  \citep{Salim16}. The typical difference ranged from $0.03 - 0.13\ \mathrm{dex}$. SFR is similarly measured using fits to the galaxy photometry taken from the SDSS and GALEX. SFRs are calculated using the method outlined in \cite{Salim07} with dust attenuation calculated using the 
\cite{CharlotFall00} model. \cite{Salim16} found that the MPA-JHU SFRs showed no evidence of bias when compared with two independent SFR measures, including mid-IR SFRs. \\

Whilst the MPA-JHU catalogue is formally deprecated by the SDSS we found the three alternative catalogues (Wisconsin, Portsmouth and Granada\footnote{For more information about these catalogues, visit the \href{https://www.sdss.org/dr16/spectro/galaxy/}{\textcolor{blue}{SDSS galaxy properties page}}}) were insufficient for our purposes. The Wisconsin catalogue did not calculate any SFR values, nor did they present any spectral quantity from which one could be derived. The Portsmouth catalogue does calculate the appropriate quantities but 64\% of their SFRs have a zero value across a wide range of magnitudes. In addition, the non-zero SFRs are gridded, implying a lack precision at lower values. The Granada catalogue has the most promising quantities, having a greater proportion of non-zero SFRs. However, there is also evidence of gridding at lower values. Furthermore, there appears to be a strict limit applied to the specific SFR. It is unclear why this limit has been applied and it has the effect of removing the galactic main sequence of star formation. MPA-JHU is not subject to any of the above issues. It is a widely used catalogue with robust stellar mass and SFR values. Thus, despite its deprecated status, MPA-JHU is the best catalogue for this work and its quantities are used throughout.\\

The X-ray data come from the 3XMM DR7 catalogue \citep{Rosen16}. It is based on 9,710 pointed observations with the XMM-Newton EPIC cameras which have a field of view $\approx 30$' and cover the energy range $\sim 0.2 -12$ keV.  DR7 contains $\sim 400,000$ unique X-ray sources based on 727,790 individual detections. For our study we use the unique source list rather than the individual detections. Our results are thus averaged over several individual observations for a significant number of sources. We summed fluxes in the 2 - 4.5 keV and 4.5 - 12 keV bands and converted them to luminosities in the 2 - 12 keV energy range using the MPA-JHU redshifts, assuming $\Gamma = 1.7$. We continue to use this catalogue, instead of 4XMM, because our analysis requires the use of comprehensive upper limits data accessible through Flix\footnote{Found at \href{https://www.ledas.ac.uk/flix/flix_dr7.html}{https://www.ledas.ac.uk/flix/flix\_dr7.html}} \citep{Carrera07}. 3XMM DR7 is the most recent version of the serendipitous sky survey available with this infrastructure.

\subsection{Position Matching}
\label{sec:matching}
In \cite{Birchall20} the dwarf galaxies in the MPA-JHU catalogue were matched to 3XMM with a simple but robust statistical technique. The small sample size allowed for individual inspection of degenerate matches i.e. galaxies matched to multiple X-ray signals, and vice versa. By removing the mass threshold, we are massively expanding our potential sample. This opened up the possibility of thousands of potential degenerate matches between MPA-JHU and 3XMM. An automated method that assessed the strength of these degenerate matches would be needed to feasibly process such a large catalogue. \\

To overcome this challenge we turned to the ARCHES cross-correlation tool, \texttt{xmatch} \citep{Pineau17}. It is an astronomical matching tool able to identify the counterparts of one catalogue to multiple others, whilst also computing probabilities of associations using background sources and positional errors. In consultation with its creator, we wrote a script to match MPA-JHU and 3XMM DR7. It uses a standard approach adopted when working with XMM, where the data and matching process is broken down into individual XMM fields. Every 3XMM and SDSS object within 15' of the field's central co-ordinates was found. Once complete, a level of quality control was applied, to isolate the most secure X-ray detections - those entries whose absolute position error was no greater than 4" and whose physical extent was less than 10". Crucially, \texttt{xmatch} calculated the Bayesian probability of association and non-association for each remaining pair of X-ray and galaxy co-ordinates within the current 3XMM field. After repeating this process for all 3XMM fields, \texttt{xmatch} identified 110,274 potential matches.\\

Unfortunately, by focusing on individual fields the matching process did account not for any overlap that might occur between them. Before we assessed the match strengths, we removed repeating galaxy or X-ray matches produced from overlapping fields. Once done, we isolated the SDSS objects with an entry in MPA-JHU. All the SDSS co-ordinates were matched to MPA-JHU co-ordinates within 1", leaving 3,357 galaxies with mass and SFR estimates. Next, we identified the most likely X-ray \& galaxy pairs. We chose to set a 90\% probability of association as the matching threshold which produced a well-matched sample of 1,559 X-ray emitting galaxies. 

\subsection{Clarifying the Effect of High Optical Emission on the SFRs \& Stellar Masses}
\label{sec:HOC_AGN}

\begin{figure} 
\includegraphics[width=\columnwidth]{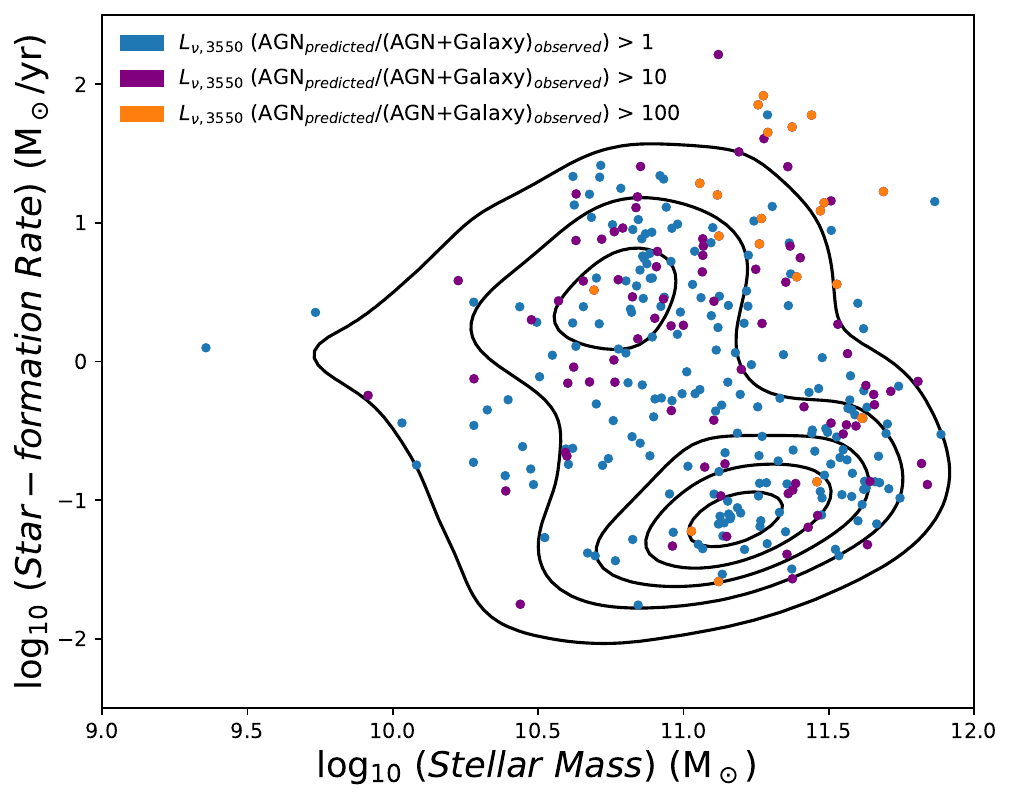}
\centering
\caption{Comparing the stellar mass and SFR of galaxies with a high, predicted optical AGN contribution ($L_{\nu, 3550\dot{A}} (\mathrm{AGN_{predicted}/(AGN + Galaxy)_{observed}})$; coloured points) against the the full X-ray emitting galaxy sample (black contours). The contours encompass 90\%, 70\%, 50\%, 30\% and 10\% of the galaxy sample. See section \ref{sec:HOC_AGN} for more information on how $L_{\nu, 3550\dot{A}} (\mathrm{AGN_{predicted}/(AGN + Galaxy)_{observed}})$ was calculated.}
\label{fig:Mass_SFR_Check}
\end{figure}

MPA-JHU stellar masses were calculated by fitting each galaxy's spectral energy distribution (SED) \citep{Kauffmann03} using optical data from the SDSS; SFRs use the optical SEDs and add UV data \citep{Salim07} from GALEX \citep{Martin05} as well. As outlined above, we were confident that these values are robust and well-suited to our study. However, we expected to find high luminosity AGN in this sample. Thus we wanted to check that strong central optical emission, likely originating from an AGN, was not biasing these quantities within our sample of candidate AGN-hosting galaxies. To determine what proportion of the total optical emission originated from the AGN, we drew upon techniques previously used in \cite{Birchall20}. This technique aimed to calculate and compare the luminosity densities at $3550 \dot{A}$ originating from the AGN and compare to it the overall optical emission ($L_{\nu, 3550\dot{A}} (\mathrm{AGN_{predicted}/(AGN + Galaxy)_{observed}})$). \\

For this sake of this calculation we assumed that every galaxy in this sample hosted an AGN. To predict the AGN's contribution to the total observed optical light we used the X-ray to optical spectral index published in \cite{LussoRisaliti16}. Using the X-ray emission ensured there was little contamination from stellar processes. The \cite{LussoRisaliti16} relation has the form, 
\begin{equation}
\log_{10}\ (L_{\nu,\ 2500 \dot{A}}) = \frac{1}{0.6}\  (\log_{10}\ (L_{\mathrm{\nu,\ 2\ keV}}) - 7)
\label{eq:LS16}
\end{equation}\\
It relates the luminosity density at 2 keV ($L_{\mathrm{\nu,\ 2\ keV}}$) to that at $2500\dot{\mathrm{A}}$ ($L_{\nu,\ 2500\dot{\mathrm{A}}}$) from a sample of SDSS quasars. Assuming our galaxies follow the same relation, it can provide a useful starting point in estimating the effects of the AGN on optical observations. 
To find the luminosity density at 2 keV, we calculated the geometric means of 3XMM bands 2 \& 3 and bands 4 \& 5, giving the luminosity densities at 1 keV and 5 keV respectively. Using linear interpolation between these values we calculated $L_{\mathrm{\nu,\ 2\ keV}}$ and used it in equation \eqref{eq:LS16} to find $L_{\nu,\ 2500\dot{\mathrm{A}}}$. Translating $L_{\nu, 2500\dot{A}}$ to the required emission at 3550$\dot{\mathrm{A}}$ required the composite UV-optical quasar spectrum from \cite{VandenBerk01}. They model this region as a power law spectrum, $f_\mathrm{\nu} = c \nu^{-\alpha}$, where $\alpha = 0.44$. From this an average spectrum was constructed and the flux density at $3550\dot{\mathrm{A}}$ extracted.\\

The total observed optical emission was calculated by converting the SDSS U-band magnitude to a flux density at $3550 \dot{\mathrm{A}}$. We then divided the predicted AGN emission at $3550 \dot{\mathrm{A}}$ by the total optical emission to obtain the $L_{\nu, 3550\dot{A}} (\mathrm{AGN_{predicted}/(AGN + Galaxy)_{observed}})$ ratio. We plotted the stellar mass and SFR distribution of the 1,559 X-ray emitting galaxies as black contours in figure \ref{fig:Mass_SFR_Check}. Overlaid on this figure are coloured points representing galaxies with high predicted AGN contributions.\\

The \cite{LussoRisaliti16} relation assumes the AGN are un-obscured. Since MPA-JHU selects only narrow-line objects this predicted AGN luminosity will be an upper limit on its contribution. Thus some of these AGN have predicted optical contributions that are greater than the actual observed values. However, these sources constitute less than $20\%$ of the galaxy sample. This fraction drops to $\sim 7\%$ at $L_{\nu, 3550\dot{A}} (\mathrm{AGN_{predicted}/(AGN + Galaxy)_{observed}}) > 10$ and $\sim 1\%$ at $L_{\nu, 3550\dot{A}} (\mathrm{AGN_{predicted}/(AGN + Galaxy)_{observed}}) > 100$. 
Figure \ref{fig:Mass_SFR_Check} clearly shows that even for these sources, with a strong predicted AGN contribution to the optical light, the stellar mass and SFR values calculated are consistent with the wider X-ray emitting galaxy distribution. Thus we continued to confidently use these values.

\subsection{Identifying AGN}
\label{sec:AGNcriterion}

\begin{figure*}
    \centering
    \includegraphics[width=.8\paperwidth]{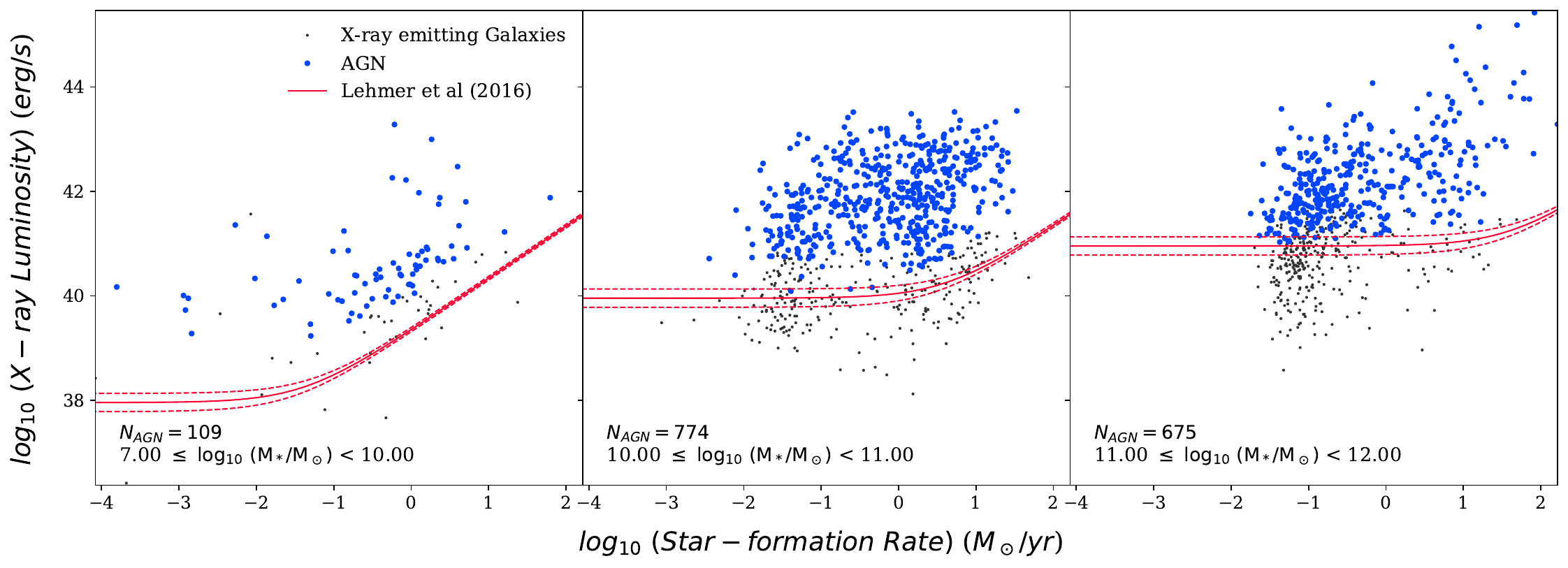}
    \caption{Observed X-ray luminosity against SFR for the 1,559 X-ray emitting galaxies which are candidates to host AGN. For clarity, these candidates were split up into three panels of different host galaxy masses (highlight in text on each panel). Objects that met or exceeded the threshold are plotted in blue, those that did not are grey. The red line is the \protect \cite{Lehmer16} equation, the dominant source of X-ray emission prediction, described in section \ref{sec:AGNcriterion}. Variation due to uncertainty in coefficients is highlighted by the dashed lines. By comparing it across plots we can clearly see the variation in mass and SFR dependencies.}
    \label{fig:Lx_SFR_Mass}
\end{figure*}

To identify emission from an AGN, we modelled the combined emission coming from other X-ray emitting sources - X-ray binary stars and hot gas emission - and compared it to the observed X-ray luminosity, following the process used in \cite{Birchall20}. \\

X-ray binary star contributions, $L_{\mathrm{XRB}}$, were modelled using the relationship provided in \cite{Lehmer16} (see also \cite{Aird17}), 
\begin{equation}
L_{\mathrm{XRB}} = \alpha_0 (1+z)^\gamma M_\mathrm{*} (\mathrm{M_\odot}) + \beta_0 (1+z)^\delta SFR (\mathrm{M_\odot yr^{-1}})
\label{eq:Lehmer2016}
\end{equation}\\
where $\log_{10} (\alpha_0) = 29.37 \pm 0.15$, $\gamma = 2.03 \pm 0.60$, $\log_{10} (\beta_0)=39.28 \pm  0.03$ and $\delta = 1.31 \pm 0.13$ for 2 - 10 keV. The low-mass X-ray binary (LMXB) contribution is correlated to the stellar mass ($M_*$), the high mass X-ray binary (HMXB) contribution to the  SFR and the redshift ($z$) dependence accounts for changes in metallicity and evolution of the XRB population. \\

Hot gas in the interstellar medium could also account for some portion of the observed X-ray emission. Its contribution, $L_{\mathrm{Gas}}$, can be estimated using the \cite{Mineo12b} relation,
\begin{equation}
L_{\mathrm{Gas}} = (8.3 \pm 0.1) \times 10^{38}\ SFR\ (\mathrm{M_\odot\ yr^{- 1}})
\end{equation}\\
We calculated the expected emission from the hot gas using \cite{Mineo12b} and added this to their $L_{\mathrm{XRB}}$. Despite the relatively low magnitude of this relationship, it was still important to calculate as all significant alternative X-ray sources needed to be considered. \\

Once both these contributions had been calculated, we summed them and compared this quantity to the observed X-ray luminosity, $L_{X,Obs}$. We considered any object which met or exceeded the following criterion to be an AGN, 
\begin{equation}
\label{eq:excessCriterion}
\frac{L_{\mathrm{X,Obs}}}{L_{\mathrm{XRB}} + L_{\mathrm{Gas}}} \geq 3
\end{equation}\\
Based on this criterion, 949 X-ray emitting galaxies were classified as AGN. Figure \ref{fig:Lx_SFR_Mass} shows a breakdown of this classification process, comparing the observed X-ray luminosity with the SFR for every X-ray emitting galaxy and highlighting whether it has been classified as an AGN. The sample was split into several mass bins for clarity but it also highlights the changing nature of the \cite{Lehmer16} equation. Each illustrated line was calculated using the mean stellar mass, indicated on the panel, and mean redshift of the bin. So each line should be considered an illustrative threshold. As the stellar mass increases, so too does the normalisation of equation \eqref{eq:Lehmer2016} at low SFR. So for the highest mass galaxies, this relationship behaves almost like a luminosity cut of $>10^{41}$ erg/s, lower than the typical threshold of $10^{42}$ erg/s \citep{BrandtAlexander15}. At such low redshifts and SFRs, however, we trust that the selection criterion will continue to identify AGN activity. This increase in normalisation is believed to originate from a growing population of long-lived low mass X-ray binaries that occurs in more massive galaxies. Variation due to uncertainty in coefficients is highlighted by the dashed lines. \\

\begin{figure}
    \centering
    \includegraphics[width=\columnwidth]{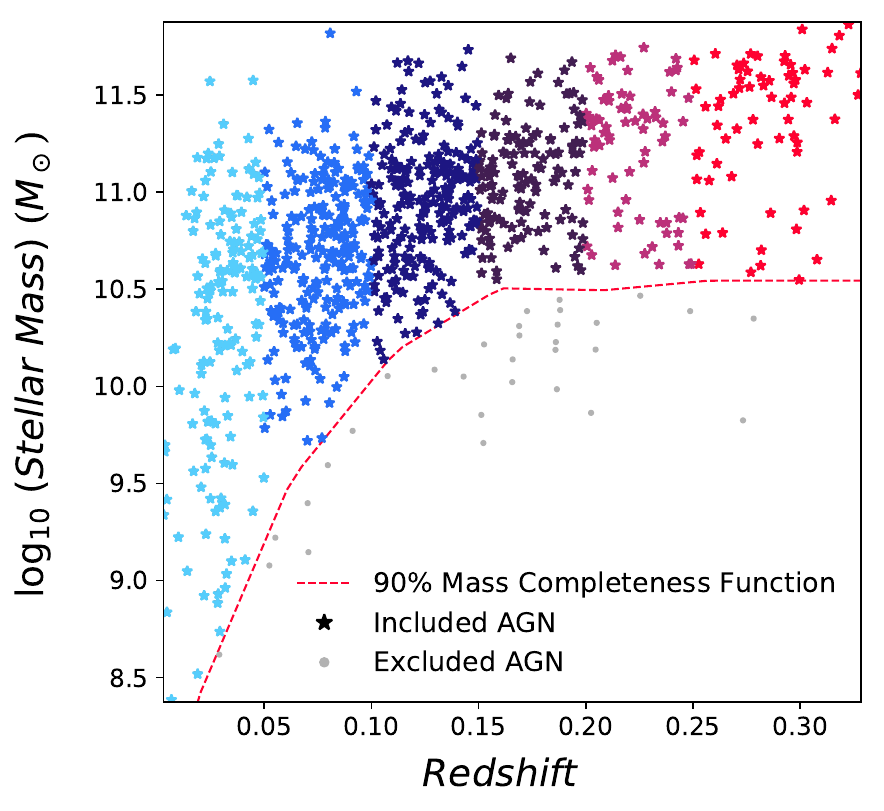}
    \caption{Redshift against stellar mass for the 949 AGN identified with equation \eqref{eq:excessCriterion}. This sample has been limited by the 90\% mass completeness function (red, dashed line), resulting in the removal of 32 AGN (grey points). The remaining 917 AGN have been split into 6 redshift bins outlined in section \ref{sec:AGNcriterion}, and separated by colour for clarity.}
    \label{fig:Mass_Z_Limits}
\end{figure}

To measure an accurate AGN fraction for this sample we had to ensure it was a statistically complete sample of galaxies above a given stellar mass limit. Thus, we took galaxies of all masses in narrow bands of redshifts from MPA-JHU and determined the mass range which contained $\sim 90\%$ of the galaxies. This process allowed us to create a 90\% mass completeness function. Figure \ref{fig:Mass_Z_Limits} shows the distribution of AGN on the redshift and stellar mass plane, with the completeness function highlighted as a red line. We have split this into 6 redshift intervals between 0 and 0.35 - 5 of width 0.05 and the final being 0.1 in width. These bins are highlighted by their colour in the figure. Only those AGN which lie above the completeness function were binned, resulting in the removal of 32 objects.  \\

\section{BPT Classification}
\label{sec:BPT}
AGN activity can impact the host galaxy's emission across the electromagnetic spectrum. The BPT diagnostic \citep{BaldwinPhillipsTerlevich81} is a commonly used technique to identify the primary source of ionising radiation in the optical part of the spectrum. By comparing the ratios of various emission lines, we can gain insight into whether star formation, AGN or a composite of both processes dominate in any given galaxy. In \cite{Birchall20}, we found that this diagnostic missed around 85\% of our X-ray selected AGN in dwarf galaxies. 
Now with this larger sample we were able to investigate how the accuracy of this diagnostic changed with stellar mass.\\ 

\begin{figure*} 
\includegraphics[width=.8\paperwidth]{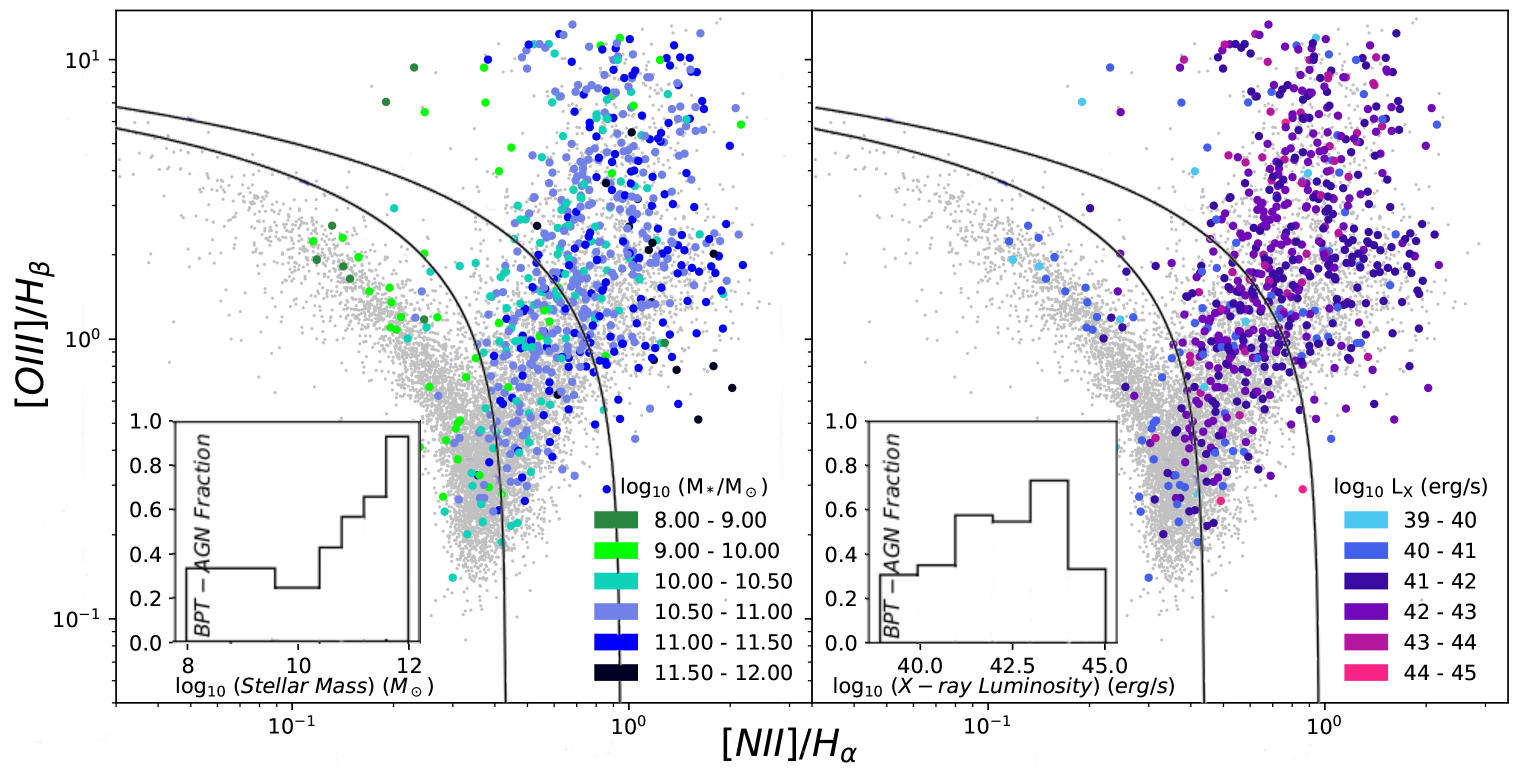}
\caption{BPT diagnostics for the 658 X-ray selected AGN with significant detections in all the emission lines shown. In the left-hand panel, the colour of each AGN point represents the mass of the host galaxy, outlined in the legend. In the right-hand panel, the colour of each AGN point represents the observed X-ray luminosity of the AGN. Inset into each panel, we see the fraction of X-ray selected AGN with the same classification from the BPT diagnostic, as a function of the relevant observable.}
\label{fig:BPT}
\end{figure*}

Of the 917 AGN hosts we identified using X-ray selection techniques, 658 had significant detections ($\frac{\mathrm{Line\ Flux}}{\mathrm{Line\ Flux\ Error}} > 3$) in each of the required emission lines so these were used in our analysis. In the left-hand panel of figure \ref{fig:BPT}, we show the results of the BPT analysis with our AGN sample plotted as large points, coloured to indicate their stellar mass. Black lines separate the AGN hosts into different classifications: objects with ionisation signatures predominately from AGN lie in the top-right, those dominated by star formation in the bottom-left, and those that have composite spectra are in the central region 
\citep{Kewley01, Kauffmann03b}. Underneath these points, in grey, are a subset of the MPA-JHU galaxies used to illustrate the underlying BPT distribution. Encouragingly, the vast majority of our AGN appear to lie within the AGN region. However, the colours clearly show that most of the objects found here are higher mass galaxies, with the lower mass points favouring the star-forming and composite regions. The behaviour exhibited by these low mass galaxies mirrors what we found in \cite{Birchall20} and is consistent with a growing body of work suggesting that these optical spectroscopic measurements are less effective at identifying AGN in lower mass galaxies \citep{Agostino19,Cann19}. \\

To further interrogate this trend, we calculated the proportion of X-ray-selected AGN with the same BPT classification as a function of stellar mass. Inset into the left-hand panel of figure \ref{fig:BPT}, we can see that the effectiveness of the BPT diagnostic in reproducing our X-ray based classification increases with stellar mass. At the low mass end we can see similar behaviour to that found in our previous work, with the diagnostic missing around 70\% of our dwarf galaxy AGN. However, the accuracy rises steadily until it identifies 93\% of our AGN in the highest mass host galaxies. \\

Our previous paper concluded that the AGNs' relatively low luminosity caused their stark mis-classification. In the right-hand panel of figure \ref{fig:BPT} we investigate how the changing X-ray luminosity affects the BPT classification. Inset in this panel is a distribution showing the changing BPT accuracy with  X-ray luminosity. Whilst there is an increase in BPT accuracy, it clearly peaks before reaching the highest luminosity AGN: rising from 31\% accurate, in the lowest mass bin, to 73\% in the most populous. Clearly then, increasing X-ray luminosity is not the only variable that affects mass dependence of the BPT accuracy. \\

\cite{Behroozi19} collated the quenched fraction of galaxies in the nearby universe \citep[based on][]{Bauer13, Moustakas13, Muzzin13} and showed it increases strongly with stellar mass. This means higher mass galaxies are much less likely to have their emission lines driven by star formation activity. As we move towards higher mass galaxies, the typically higher luminosity AGN combined with this drop in star formation would see AGN-driven signatures dominate and produce the concurrent increase in BPT accuracy.

\section{Specific Black Hole Accretion Rate}
\label{sec:sBHAR}
AGN are powered by mass accretion onto a galaxy's central supermassive black hole however we do not gain any insight into this accretion process by only looking at the observed X-ray luminosity. For example, one black hole growing at a higher accretion rate in a lower mass galaxy could produce a similar X-ray luminosity to another black hole with a lower accretion rate in a high mass galaxy. To break this degeneracy and understand more about how the black hole activity is distributed across our observed AGN sample, we looked at the specific black hole accretion rate (sBHAR), $\lambda_\mathrm{sBHAR}$. This quantity compares the bolometric AGN luminosity of the galaxy with an estimate of the black hole's Eddington luminosity to give an indication of how efficiently the black hole is accreting. It is found using,
\begin{equation} \label{eq:sBHAR}
    \lambda_{\mathrm{sBHAR}} = \frac{25 L_{\mathrm{2-10 keV}}}{1.26 \times 10^{38} \times 0.002M_*} \approx \frac{L_{\mathrm{bol}}}{L_{\mathrm{Edd}}}
\end{equation}
and is taken from \cite{Aird12}. Our use of $0.002 M_*$ implies a correlation between the mass of the stellar bulge and SMBH which is dependent on galaxy morphology, among other properties \citep[e.g.][]{BlantonMoustakas09,Jahnke09}. However, our intention behind using this correlation is not to accurately recreate an Eddington ratio, instead it is to present an Eddington-scaled accretion rate quantity for our galaxy sample. By using this scaling to calculate sBHAR - a tracer of the rate of black hole growth relative to the host galaxy's stellar mass - we can directly compare our results to prior work at higher redshifts.\\

\begin{figure}
    \centering
    \includegraphics[width=\columnwidth]{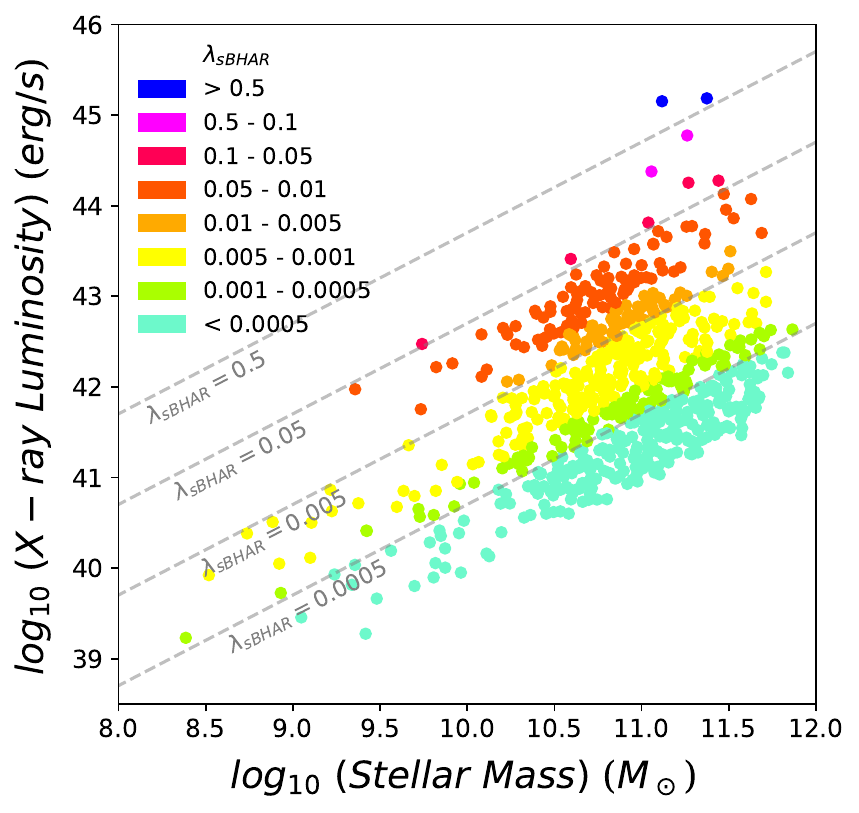}
    \caption{Stellar mass against the observed X-ray luminosity for the 917 AGN detected in section \ref{sec:AGNcriterion}. Each point has been assigned a colour to provide an indication of the accretion rate ($\lambda_\mathrm{sBHAR}$) of the galaxy's central supermassive black hole, calculated in section \ref{sec:sBHAR} . Several grey lines of constant sBHAR have also been plotted for reference.}
    \label{fig:sBHAR}
\end{figure}

Figure \ref{fig:sBHAR} shows how our observed AGN sample are distributed on the stellar mass and X-ray luminosity plane, with the data point colour giving an indication of the sBHAR of each galaxy. We have also highlighted lines of constant sBHAR to give a sense of what luminosities would be expected from a host galaxy of a given mass at those accretion rates. This figure further highlights that the vast majority of the observed AGN sample lies above stellar masses of $10^{10}\ \mathrm{M_\odot}$. There is also a clear preference for black holes with relatively low accretion rates. Most of them have accretion rates that are less than 0.5\% of their Eddington luminosity and only a handful of the most massive galaxies venturing above 10\%. 

\section{Completeness-corrected Probability Distributions}
\label{sec:Prob_Dists}
Despite using AGN selection criteria that modify the X-ray luminosity threshold depending on the mass, SFR and redshift of the host galaxy, figure \ref{fig:sBHAR} clearly shows that our sample as it is currently constructed preferentially identifies AGN found in the highest mass galaxies. In this section we will summarise and apply the methods developed in \cite{Birchall20} to reduce the effects of this bias. In doing so we will create a series of probability distributions comparing different configurations of host galaxy properties. From these distributions we aim to understand more about the underlying distribution of AGN in the nearby Universe.

\subsection{Calculating Completeness Corrections}
\label{sec:Completeness_Corrections}
Using the 3XMM serendipitous sky survey as our X-ray catalogue introduced a significant amount of variation in the detection sensitivity. Such variation brings with it the possibility that lower luminosity AGN within the overlapping region of 3XMM and MPA-JHU could have been missed as the flux limit was insufficient to detect it. To overcome this, we established the fraction of the parent galaxy sample that lie within 3XMM fields with sufficient sensitivity to detect an AGN above a given luminosity threshold. \\

To perform this analysis, we made use of Flix \citep{Carrera07}, 3XMM’s upper limits service. It provides upper limits broken down by band and instrument for the whole of 3XMM. From this data we chose the upper limits and observed fluxes from band 8 in the PN camera as it covers the entire energy range of 3XMM. Because of this restriction, we also limited the observed AGN to only those with detections in that same band. Not all objects will meet this detection threshold and some XMM fields will not have been observed by the PN camera so our observed AGN sample was reduced from 917 to 739.\\

To characterise this sensitivity variation, we needed to find the X-ray detection upper limits of all the galaxies in the region of MPA-JHU that have coverage from 3XMM. There were 28,545 MPA-JHU galaxies found within 3XMM whose co-ordinates were uploaded to Flix. It returned the X-ray flux upper limits at the co-ordinates of 25,949 galaxies – our parent sample.
We then extracted the flux upper limits for a given range of stellar masses and redshifts. 
These fluxes were then converted into luminosities using the redshift of the galaxy associated with each upper limit. Finally, these upper limits were used to construct a cumulative histogram function normalised by the total number of galaxies in the current mass and redshift range. These luminosity sensitivity functions allowed us to determine the fraction of galaxies where an AGN could have been detected above a given X-ray luminosity within a range of masses or redshifts.

\subsection{Creating the Probability Distributions}
\label{sec:Create_Prob_Dists}
We are interested in calculating how the probabilities of finding AGN vary in the nearby Universe as a function of stellar mass and redshift. To illustrate the process of constructing these probability distributions, we will focus on how the probability varies with X-ray luminosity and stellar mass.\\

\begin{figure*}
    \centering
    \includegraphics[width=.7\paperwidth]{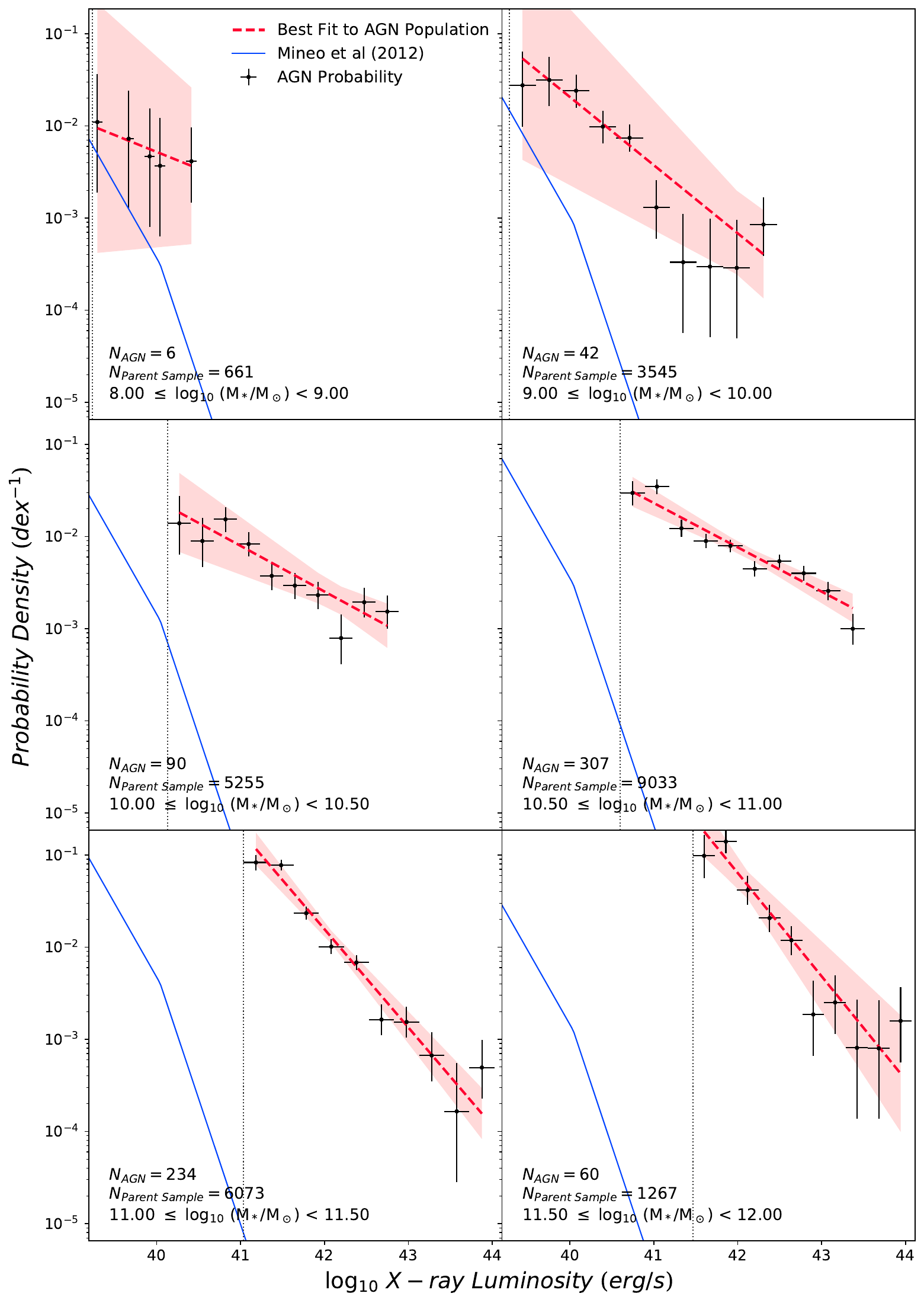}
    \caption{Probability of finding an AGN in the nearby Universe, using the completeness-corrected AGN samples, as a function of observed X-ray luminosity and stellar mass. Power laws (dashed red lines) have been fit to the data in each panel and displayed alongside their $1\sigma$ uncertainty (pale red region). Information about the number of PN-8-detected AGN and parent galaxies in the each mass range are printed on the panels. The blue line shows the average number of high luminosity stellar sources like ULXs. The form of these lines are based on the \protect \cite{Mineo12a} XLF, scaled by the average SFR of the galaxies in each bin. Dashed grey lines indicate the AGN detection limit in each bin. See section \ref{sec:Create_Prob_Dists} for more details on how these plots were constructed.}
    \label{fig:Example_Prob_Dist}
\end{figure*}

For this configuration of host galaxy properties, we split up the parent sample into a series of stellar mass bins. In each of these stellar mass bins, the data was further broken down as a function of X-ray luminosity, producing an observed AGN count distribution. In addition, we constructed a bespoke luminosity sensitivity function, as outlined in section \ref{sec:Completeness_Corrections}. Each bin's count distribution was divided by the correction fractions extracted from this sensitivity function to recover the expected number of galaxies sensitive enough to detect an AGN as a function of luminosity. Finally, we divided these corrected AGN counts by the parent sample size in this bin to produce a series of probability distributions. The results of this process are shown in figure \ref{fig:Example_Prob_Dist}. It shows the probability of finding an AGN as a function of luminosity in bins of increasing mass. Some useful reference information is printed on each panel including the sizes of the AGN and parent samples, and the stellar mass range of the galaxies included therein. \\

By applying corrections extracted from the luminosity sensitivity function to the observed AGN counts we were able to provide robust measurements of the true incidence of AGN within the nearby galaxy population.
In figure \ref{fig:Example_Prob_Dist} we see that there is an abundance of AGN across stellar mass despite the clear favouring of higher mass AGN seen in figure \ref{fig:sBHAR}. Of particular significance are the continuation of the findings from \cite{Birchall20}: AGN populations are well described by power law distributions, with AGN being found predominantly at lower X-ray luminosities \citep[see also][and others]{Aird12}. \\

To calculate the errors in each probability data point we used the confidence limits equations presented in \cite{Gehrels86} meaning that the size of the error is determined based on the number of detected AGN in a given bin. \\

One final check was performed on the AGN populations shown in figure \ref{fig:Example_Prob_Dist}. Section \ref{sec:AGNcriterion} highlights how we have considered contamination from X-ray binaries and hot gas, but there remained the possibility that this emission could have originated from an ultra-luminous X-ray source (ULX), particularly in lower mass galaxies. To check whether a ULX could account for this emission, we used the \cite{Mineo12a} X-ray luminosity function (XLF). XRB populations, including ULXs, were modelled using a two-part power law normalised by the host galaxy’s SFR. In each stellar mass bin, XLFs  associated with each individual galaxy were calculated and averaged to show how the predicted number of ULXs compared to our AGN observations. The blue lines in each panel of figure \ref{fig:Example_Prob_Dist} show that the vast majority of the data points do not overlap with the ULX XLFs. The only point at which the two distributions do meet is at the low luminosity end of the lowest stellar mass bin. \\

Since the overwhelming majority of our sample are all confirmed AGN we could confidently fit power laws to each of these panels. They have the following form,

\begin{equation}
    \label{eq:powerlaw}
       p(X) = A \left( \frac{X}{x'} \right) ^k d \log_{10}X
\end{equation}\\
where $p(X)$ is the probability of observing an AGN with corresponding X-axis quantity, $X$ and centred on a value $x’$. For this probability distribution configuration, each power law is centred on the median luminosity for the sample, $\log_{10} x’ = 42.1$ (and $\log_{10} x' = -2.55$ for sBHAR). The power laws are shown as dashed red lines in each panel and more clearly identify how the probability of finding an AGN, within the highlighted mass range, changes as a function of X-ray luminosity. The pale red regions surrounding each power law function give an indication of the error in each fit. This was calculated by performing a $\chi^2$ fit with equation \eqref{eq:powerlaw} to the black data points in each bin and their associated errors. Fit parameter errors were estimated by taking the square-root of the covariance matrix’s diagonal. With this we could outline the extent of the uncertainty in each fit. Encouragingly, in nearly all of the bins the power law fits are appropriate. \\

\subsection{Probability Distribution Comparison}
\label{sec:prob_dist_comp}
Using the method outlined in section \ref{sec:Create_Prob_Dists}, we created a number of other probability distributions to investigate how different configurations might shed light on the AGN population in the nearby Universe. As can be seen in figure \ref{fig:Prob_Dist_Comparison}, we have added redshift and sBHAR to our properties of interest.  By including redshift in our range of host galaxy properties, we can gain some insight into how the changing conditions of the Universe might affect the AGN population. Each configuration has been split up into 6 colour-coded bins of the corresponding property – either redshift (top row) or stellar mass (bottom row) – and had a probability distribution calculated from the data points in that region. For ease of comparison, each set of distributions has been placed in the appropriate column - sBHAR on the left, and X-ray luminosity on the right. The results from figure \ref{fig:Example_Prob_Dist} have been placed in the bottom-right panel of this figure. For clarity, we have only included the fits and their error region in each configuration panel. Appendix \ref{tab:Prob_Dist_Coeffs} outlines the best-fit coefficients, and associated errors, used in equation \eqref{eq:powerlaw} to create these probability distributions. In addition, the full set of data points for each configuration can be found in appendix \ref{app:Prob_Dists}. Overall, figure \ref{fig:Prob_Dist_Comparison} shows us that regardless of the host galaxy property configuration, the average number of AGN in the nearby Universe are well described by a power law. However, there are key differences between these configurations which we will outline in this section. 

\subsubsection{Slope}
In each panel of figure \ref{fig:Prob_Dist_Comparison}, there is some evidence of the slope changing with the respective property. Along the top row, we can see a series of slopes steadily steepening with redshift. This steepening effect is most pronounced in the right-hand X-ray luminosity column. However, we see in the top-left panel that AGN at higher redshifts favour more moderate accretion rates. This drives the increase in the incidence of relatively low luminosity AGN emission for the corresponding redshift bin in the top-right panel. A similar steepening can be seen in the bottom-right stellar mass, X-ray luminosity panel, however the trend is much less consistent. It shares the steepening seen in the redshift, X-ray luminosity panel but there is a lot of variation between these two extremes. 

\subsubsection{Normalisation}
\begin{figure*}
    \centering
    \includegraphics[width=.85\paperwidth]{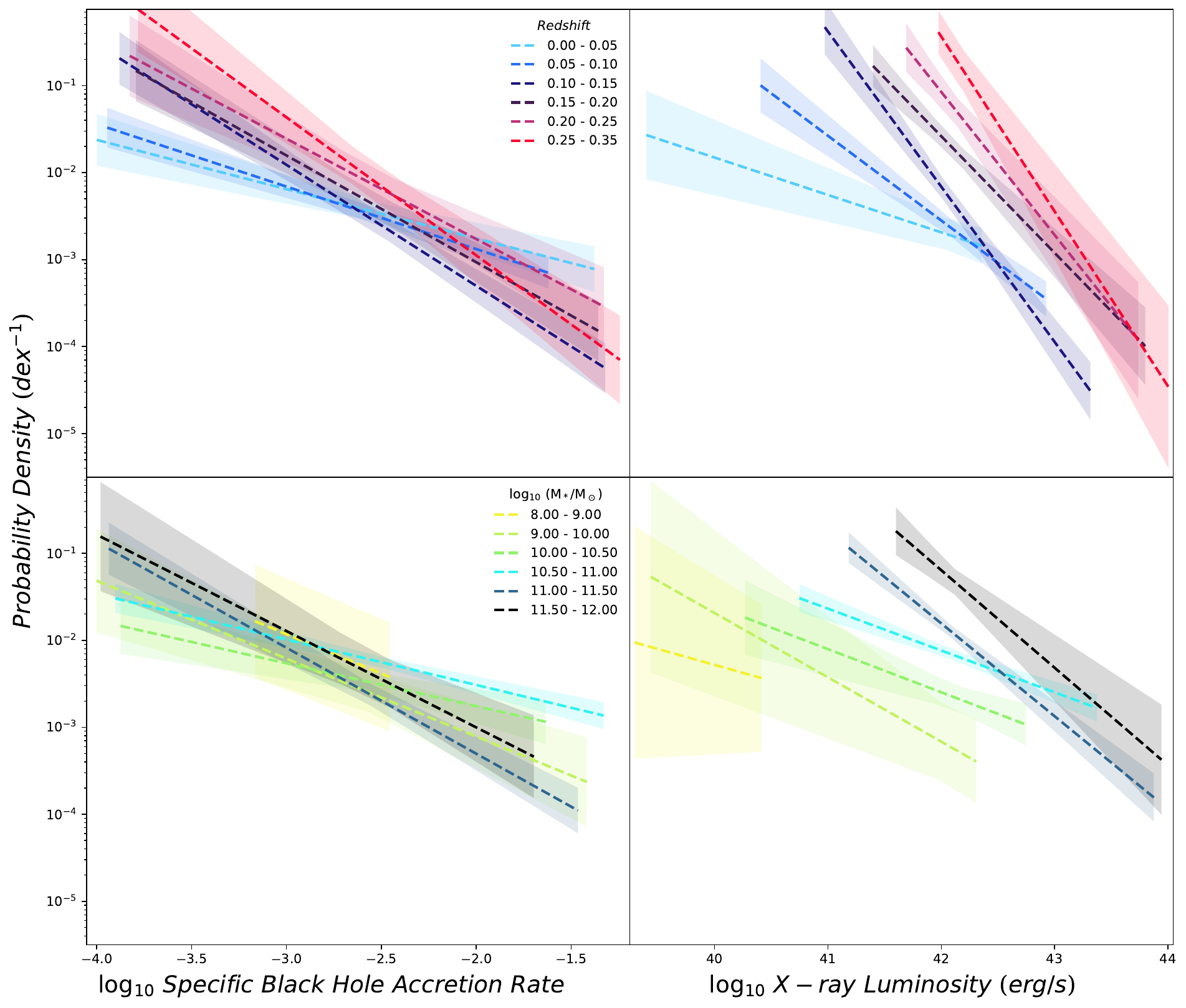}
    \caption{A comparison of the probability distributions and associated error regions calculated using different configurations of host galaxy properties. The left-hand column looks at how the probability varies with sBHAR and the right-hand column shows this with X-ray luminosity. The top row bins the AGN in redshift intervals, the bottom row does this in stellar mass. The fits for the bottom-right hand plot were taken from figure \ref{fig:Example_Prob_Dist}, the rest are presented in appendix \ref{app:Prob_Dists}.}
    \label{fig:Prob_Dist_Comparison}
\end{figure*}

As outlined in section \ref{sec:sBHAR}, studying sBHAR allows us to break observational degeneracies associated with X-ray luminosity so we can see how mass accretion is distributed across the AGN population. What is most striking about these distributions is that those in the left-hand column of figure \ref{fig:Prob_Dist_Comparison} all appear to be within the same limits. Regardless of redshift or stellar mass, there is little change in the normalisation, implying that the average amount of material a black hole accretes remains relatively consistent in the nearby Universe.\\

This trend is in stark contrast to that seen in the luminosity plots, in the right-hand column of figure \ref{fig:Prob_Dist_Comparison}. Both redshift and stellar mass distributions show consistently increasing normalisations. There are a number of processes that could be feeding into this shift: the magnitude limitation of the galaxy sample removing high redshift, low luminosity AGN; and the stellar mass and redshift dependencies of the \cite{Lehmer16} XLF in equation \eqref{eq:Lehmer2016}. However, these effects only explain why the lower limits increase and don’t account for full shifting of these luminosity-dependent distributions. This can be understood if we consider the lines of constant sBHAR in figure \ref{fig:sBHAR}: for a fixed accretion rate, stellar mass determines the range of observable luminosities. 
At higher stellar masses the much more common, lower accretion rate sources have higher observed luminosities.
When we consider that these mass and redshift bins have the same accretion limits, it is clear that changing stellar mass is the primary driver of this shift in normalisation, whether caused by explicit binning or the effect of magnitude limitation.

\section{AGN Fractions}
\label{sec:fractions}
\begin{figure*}
    \centering
    \includegraphics[width=.8\paperwidth]{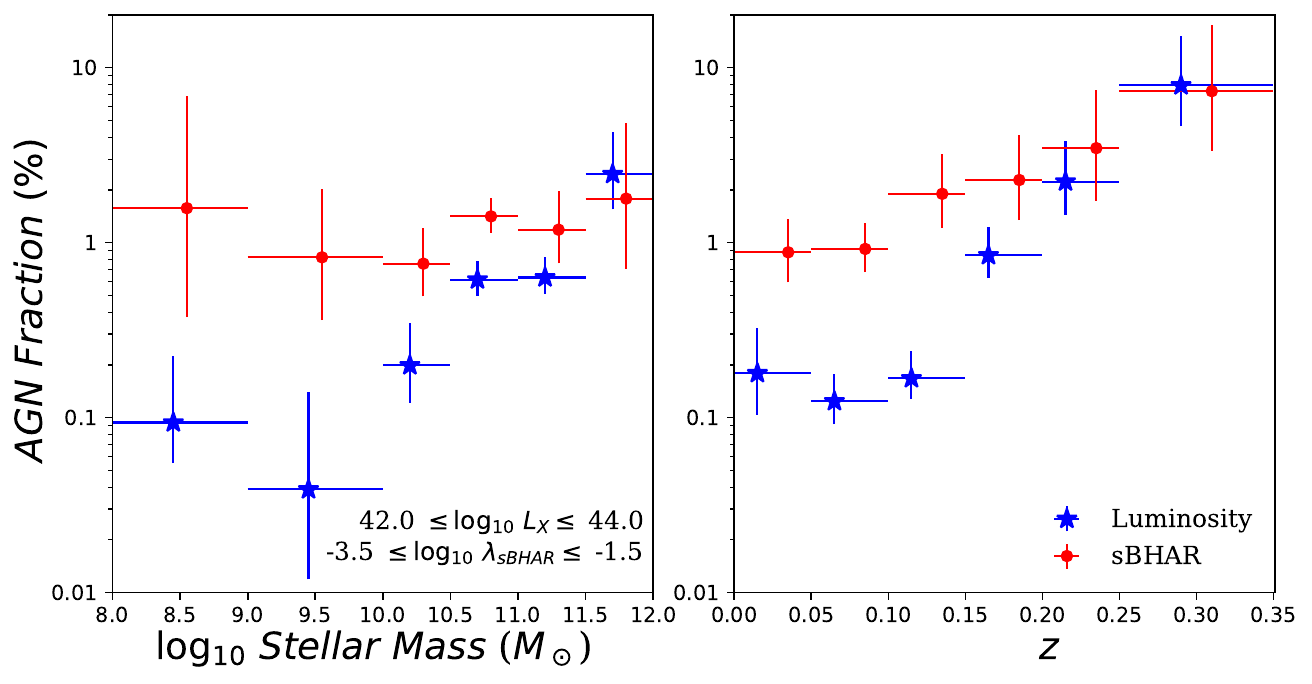}
    \caption{AGN fractions as a function of stellar mass (left-hand panel) and redshift (right-hand panel) calculated from the probability distributions in figure \ref{fig:Prob_Dist_Comparison}. Each probability distribution was integrated within the relevant limit shown in the bottom-left of the left-hand panel. There is a clear difference in AGN fraction depending on whether the luminosity (blue stars) or sBHAR (red circles) distributions were integrated.}
    \label{fig:AGN_Fraction}
\end{figure*}

Using the fits produced in section \ref{sec:prob_dist_comp}, we could calculate how the AGN fraction varied with stellar mass and redshift in the nearby Universe. The results of these calculations are shown in figure \ref{fig:AGN_Fraction}. In this section we will outline this calculation process and the significance of the results.\\

AGN fractions were calculated by integrating under the probability distributions shown in figure \ref{fig:Prob_Dist_Comparison}. Each distribution produces a single point in the corresponding position, colour and panel of figure \ref{fig:AGN_Fraction}. Upper and lower limits on each AGN fraction were calculated by integrating both edges of the error region. The limits of these integrations are shown in the bottom-right-hand corner of the stellar mass panel. Consistent limits were chosen to encapsulate the region within which the majority of the distributions are defined and to make sure each point is directly comparable. \\

This approach leaves us with two sets of AGN fractions in each galaxy property panel: luminosity-derived fractions (blue stars) and sBHAR-derived fractions (red dots). In effect, we have created two separate AGN definitions to compare. The luminosity-derived points create an observational-style definition whereby an AGN is an object emitting at an X-ray luminosity $\geq 10^{42}$ erg/s. Whereas the sBHAR-derived points define an AGN as a black hole accreting at rates $\lambda_\mathrm{sBHAR} \geq 10^{-3.5}$. As is evident from figure \ref{fig:AGN_Fraction}, these differing definitions have implications for the predicted distribution of AGN. \\

In the left-hand panel, we can see how the two AGN fraction definitions vary with stellar mass. The sBHAR-derived fractions predict a constant occupation of just over 1\% across the stellar mass range. Whereas the luminosity-derived fractions predict a much lower AGN fraction at low stellar mass before rising to agree with sBHAR. \cite{Georgakakis11} found a similar increase with stellar mass in their sample of X-ray selected AGN with $\mathrm{\log_{10} L_X} > 41$. \\

In the right-hand panel, we can see how both AGN fraction definitions vary with redshift. The sBHAR-derived fractions show a steady increase rising from 1\% to 10\%. The luminosity-derived fractions describe an even steeper increase from about 0.1\% to 10\%, agreeing with the sBHAR-derived data at the highest redshifts. \\

Further to the discussion at the beginning of section \ref{sec:data}, we also calculated these fractions using the Granada values. The trends in both luminosity- and sBHAR-derived AGN fractions seen in figure \ref{fig:AGN_Fraction} were also measured when using the Granada data.\\

\begin{figure*}

    \centering
    \includegraphics[width=.8\paperwidth]{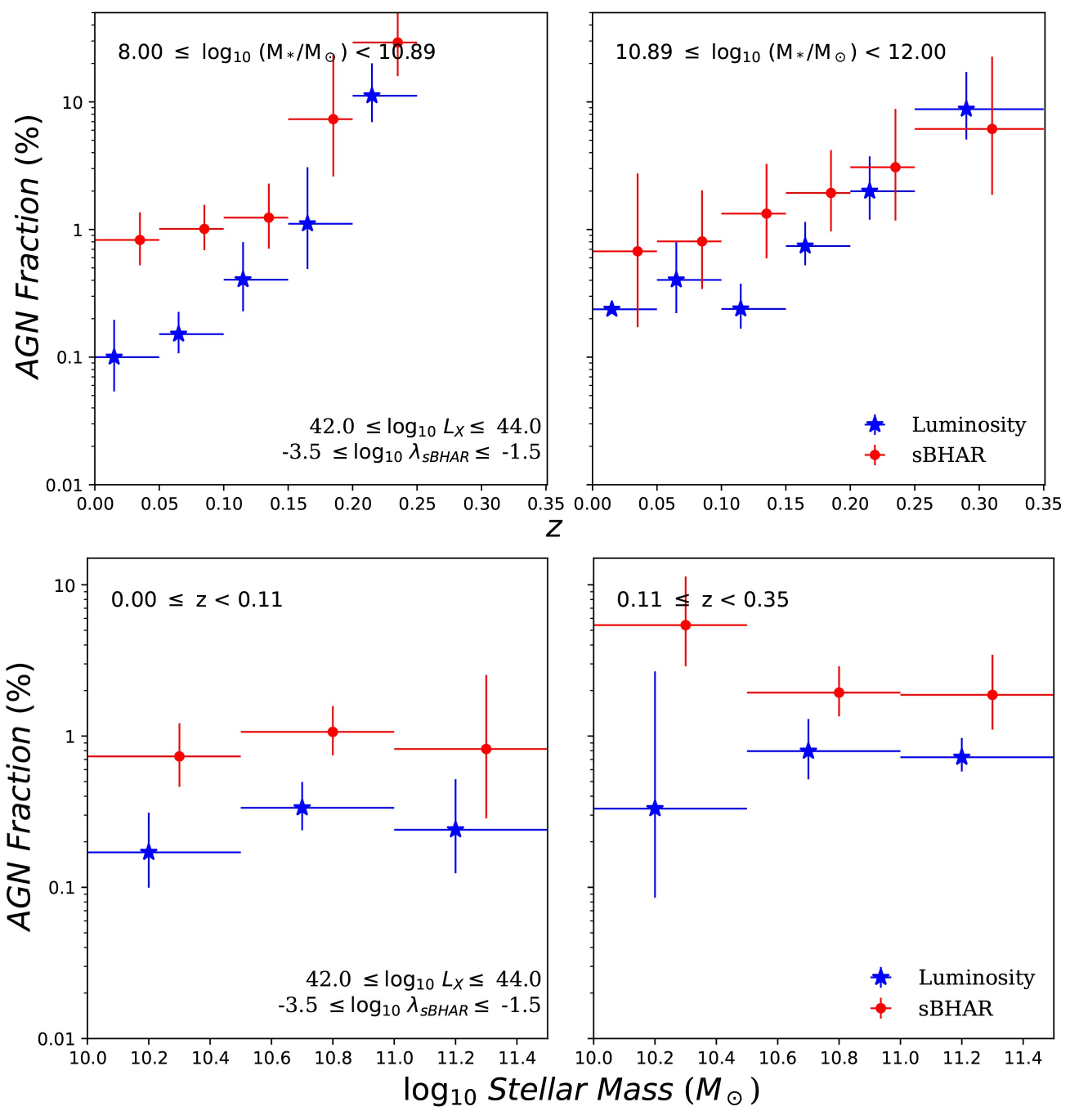}
    \caption{The top row depicts AGN fractions as a function of redshift split up into stellar mass bins - those above and below the median mass. Below are the AGN fractions as a function of stellar mass split up into redshift bins - those above and below the median redshift. The limits between which the full AGN sample has been split are shown in the top-left of each panel. As with figure \ref{fig:AGN_Fraction}, the luminosity-derived fraction are blue stars and the sBHAR-derived fractions are red circles. }
    \label{fig:Binned_Fractions}
\end{figure*}

Clearly, these two AGN fraction definitions follow different trends. Since the luminosity-derived fractions only consider higher-luminosity AGN, the sample used to calculate these will be restricted and so a drop in AGN fraction is likely. As discussed previously, all our AGN appear to be accreting over a similar range of sBHAR, so stellar mass is the main driver of changes in observed X-ray luminosity. Therefore, we would expect that lower mass AGN, which are much less likely to meet such high luminosities, are the objects most likely being excluded. In addition, figure \ref{fig:Mass_Z_Limits} shows that most of the lowest mass AGN are found at the lowest redshifts. So this would also concentrate this deficit to the lowest redshift regimes. To pinpoint where this deficit occurs, we first split the AGN by stellar mass - those above and below the median stellar mass of $10^{10.89} \mathrm{M_\odot}$ - and recalculated the AGN fractions as a function of redshift. The top row of figure \ref{fig:Binned_Fractions} confirms our expectations by showing that this luminosity-derived fraction deficit very clearly occurs in the lowest stellar mass and redshift galaxies. The right-hand, high-mass panel shows that when both limits encapsulate the majority of the AGN sample, they predict very similar trends. Thus, the luminosity-derived fractions are still subject to observational biases despite being calculated from completeness-corrected data. Other studies, with different samples, have observed this effect as well  \cite[e.g.][]{Aird12,Bongiorno16, Weigel17}. \\

Whilst the sBHAR-derived points are not free from these biases, selecting AGN based on an Eddington ratio limit has been shown to yield a wider range of AGN and host galaxy properties compared to a luminosity limit and so better represents the underlying AGN population \citep{Jones17}. It is for these reasons that we focus on the sBHAR-derived points when drawing conclusions about the underlying AGN population in the nearby Universe. The left-hand panel of figure \ref{fig:AGN_Fraction} showed us that sBHAR-derived AGN fraction was constant across the stellar mass range. Furthermore, the bottom row of figure \ref{fig:Binned_Fractions} shows that when we split the AGN by redshift - those above and below the median redshift of 0.11 - and recalculate the AGN fractions as a function of stellar mass, that this flat distribution remains. Thus, the sBHAR-derived AGN fraction does not depend on stellar mass in the nearby universe. \\

\begin{figure*}

    \centering
    \includegraphics[width=1.33\columnwidth]{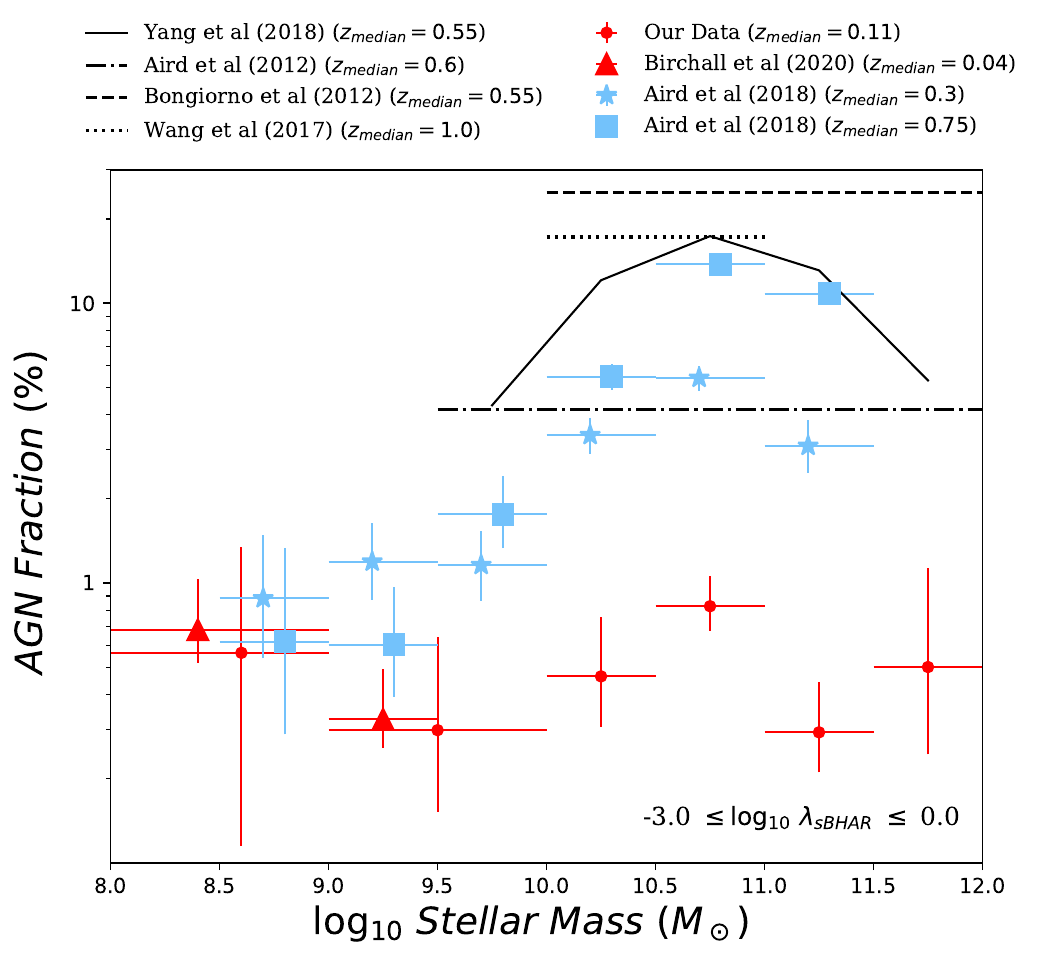}
    \caption{A comparison of how AGN fraction changes as a function of stellar mass at different redshifts. Our data (red circles) is calculated from the sBHAR-dependent fits. Alongside this are numerous other AGN fractions calculated from Eddington ratio based probability distributions within the limits shown in the bottom-right of the plot. The median sample redshift is highlighted in the legend.}
    \label{fig:Fraction_Comparison}
\end{figure*}

Figure \ref{fig:Fraction_Comparison} compares our sBHAR-derived AGN fractions with other predictions derived from Eddington ratio-based probability distributions: \cite{Aird12, Bongiorno12, Wang17, Aird18, Yang18, Birchall20}. We have increased the limits of integration compared to figure \ref{fig:AGN_Fraction} to incorporate the generally higher threshold of AGN activity used in these other studies. By shifting the lower limit from $\log_{10} \lambda_\mathrm{sBHAR} = -3.5$ to $-3$, we have halved the average AGN fraction predicted from our data to $0.5\%$. This shows that there is a significant amount of low level accretion in the nearby Universe.\\

Our results provide one of the first, robust measurements of the incidence of AGN within the nearby galaxy population. To highlight this we presented the median redshift of each paper's sample in the legend. Overall we see that as redshift increases so too does the AGN fraction. It is thought that the increased availability of cold gas at earlier times \citep{Mullaney12, Popping12, Vito14} would drive this increase in activity. Other AGN samples taken from a range of redshifts and wavelengths appear to show similar patterns of activity \citep{Aird18,Delvecchio18,Yang18}. 
To check whether our results are consistent with this evolution, we re-integrated equation (12) of \cite{Aird12} between the same sBHAR limits shown in figure \ref{fig:Fraction_Comparison}. However, this time we modified the redshift-dependent normalisation term by evaluating it at $z = 0.11$, the median redshift of our work. This produced a new fraction of $1.1\%$, consistent with most of our fractions.\\

However, figure \ref{fig:Fraction_Comparison} shows that this evolution is suppressed in lower mass galaxies. We can see significant alignment between points accounting for AGN samples measured out to $z = 1$ in the dwarf galaxy mass range ($\log_{10} \mathrm{M_*/M_\odot} \leq 9.5$); consisting of our results, and those from \cite{Birchall20} and \cite{Aird18}. It is thought that the since the potential wells for lower mass galaxies are shallower, gas particles are less likely to fall into the galactic centre and fuel the SMBH \citep{Bellovary13}. A similar drop in AGN activity at lower stellar masses has also been observed in different AGN samples \citep{Yang17, Aird18}. 

\section{Summary \& Conclusions}

This study has rigorously matched a sample of local galaxies from MPA-JHU to their X-ray counterparts in 3XMM using \texttt{xmatch}. We re-affirmed that MPA-JHU was the catalogue most suited for this task. Then, before identifying AGN, we confirmed that strong nuclear optical emission from our matched sample did not bias the stellar mass and star-formation rate measurements. With this confirmation we could use these quantities to confidently predict the X-ray luminosity due to emission from X-ray binaries and hot gas. Any galaxies that had observed X-ray emission at least three times larger than the sum of these predictions was considered an AGN host. Finally, any AGN that fell outside of the 90\% mass completeness function were removed from the sample. This left us with 917 AGN. \\

We then performed BPT analysis on the 658 AGN with significant detections in all relevant emission lines. We have demonstrated that there is a strong stellar mass dependence in the BPT diagnostic. Higher mass X-ray selected AGN are much more likely to be assigned the same classification by the BPT diagnostic. In contrast, we find little dependence in the BPT-selected fraction of AGN on X-ray luminosity. Since the quenched fraction of galaxies is believed to increase with stellar mass, then the influence of star-formation on these emission lines would drop significantly and produce the increase in BPT accuracy. 
\\

Next we investigated the activity of the black holes powering our AGN by calculating their specific accretion rates. We found that there is a clear preference for black holes with relatively low accretion rates. 
Most of them have accretion rates that are less than 0.5\% of their Eddington luminosity (assuming a nominal scaling between black hole mass and galaxy stellar mass), with only a handful of the most massive galaxies venturing above 10\%. Only two AGN have an accretion rate that is greater than 50\% of their Eddington luminosity.\\

We then corrected for observational bias in our AGN sample. Drawing on the method used in \cite{Birchall20}, we used Flix to determine the flux limits of the 25,949 MPA-JHU galaxies found within 3XMM - our parent sample. These upper limits form the basis from which a sensitivity function can be created and used to correct our observed distributions of AGN. From this we created a series of probability distributions splitting the AGN population by stellar mass and redshift, and looking at trends with observed X-ray luminosity and specific black hole accretion rate. These distributions are well-described by power laws but when compared show distinct trends. As redshift increases, the power laws steepen suggesting more moderate accretion rates and thus luminosities, are favoured at earlier times. A similar steepening trend can be seen with stellar mass, however, the progression is less certain. We also find our galaxies contain AGN accreting over approximately the same, broad range of specific accretion rates ($-3.5\lesssim \log \lambda_\mathrm{sBHAR} \lesssim -1.5$) 
regardless of stellar mass or redshift. \\

Finally, integrating under each of these distributions allows us to calculate robust AGN fractions and determine how these vary with host galaxy stellar mass and redshift. Since we have investigated the AGN distribution as a function of both X-ray luminosity and specific black hole accretion rate, we adopt two definitions of an AGN when determining the fraction. The first is pseudo-observational, defining an AGN an object with X-ray luminosity $\geq 10^{42}$ erg/s; the second defines an AGN as a black hole accreting at rates $\lambda_\mathrm{sBHAR}\geq 10^{-3.5}$. We find that the pseudo-observational luminosity-limited definition reproduces observational biases inherent in previous AGN studies so we focus on the accretion rate derived fractions. 
We find no evidence that our accretion-rate-derived AGN fractions depend on the stellar mass, finding a constant fraction of just over 1\% for $8\lesssim \log M_\mathrm{*}/M_\odot \lesssim 12$.
In addition, we find that AGN fraction increases with redshift, rising from 1\% to 10\% at $z = 0.33$. \\

We then derived AGN fractions from sBHAR distributions produced by other studies and compared them to our results. Our low redshift results help confirm the previously identified, strong redshift evolution of the AGN fraction in galaxies with $M_*\gtrsim10^{10}M_\odot$ 
brought about by an increase in the availability of cold gas. 
In contrast, our results confirm a lack of evolution at low stellar masses, indicating a constant AGN fraction out to $z=1$ \citep[see also][]{Birchall20}
The suppression in this region is thought to be driven by the shallower potential wells in lower mass galaxies that are unable to capture as much gas as their higher mass counterparts.\\

In conclusion, we have shown that AGN with a broad range of accretion rates are found across the local galaxy population. This work shows that lower mass AGN are more likely to be missed by optical selection but that careful analysis of X-ray observations will help reveal them. Our measurements quantify the AGN fraction in the local Universe and provide crucial limits on their evolution over cosmic time. \\ 

\section*{Acknowledgements}
KB acknowledges funding from a STFC PhD studentship.
JA acknowledges support from an STFC Ernest Rutherford Fellowship (grant code: ST/P004172/1) and a UKRI Future Leaders Fellowship (grant code: MR/T020989/1).

This research has made use of data obtained from the 3XMM XMM-Newton serendipitous source catalogue compiled by the 10 institutes of the XMM-Newton Survey Science Centre selected by ESA. In addition, this research made use of Astropy,\footnote{http://www.astropy.org} a community-developed core Python package for Astronomy \citep{astropy:2013, astropy:2018}. 
Funding for SDSS-III has been provided by the Alfred P. Sloan Foundation, the Participating Institutions, the National Science Foundation, and the U.S. Department of Energy Office of Science. The SDSS-III web site is \href{http://www.sdss3.org/}{http://www.sdss3.org/}.

KB would also like to thank F-X Pineau for his help with constructing the \texttt{xmatch} scripts, and Duncan Law-Green for the help accessing the Flix archive. 

\section*{Data Availability Statement}
The data underlying this article has been derived from publicly available datasets: the MPA-JHU Catalogue (based on SDSS DR8) \& 3XMM DR7. Section \ref{sec:data} outlines where these catalogues are available and how the final sample was derived. The underlying data will also be shared on request to the corresponding author.

\bibliographystyle{mnras}
\bibliography{Bibliography} 

\begin{thebibliography}{}
\makeatletter
\relax
\def\mn@urlcharsother{\let\do\@makeother \do\$\do\&\do\#\do\^\do\_\do\%\do\~}
\def\mn@doi{\begingroup\mn@urlcharsother \@ifnextchar [ {\mn@doi@}
  {\mn@doi@[]}}
\def\mn@doi@[#1]#2{\def\@tempa{#1}\ifx\@tempa\@empty \href
  {http://dx.doi.org/#2} {doi:#2}\else \href {http://dx.doi.org/#2} {#1}\fi
  \endgroup}
\def\mn@eprint#1#2{\mn@eprint@#1:#2::\@nil}
\def\mn@eprint@arXiv#1{\href {http://arxiv.org/abs/#1} {{\tt arXiv:#1}}}
\def\mn@eprint@dblp#1{\href {http://dblp.uni-trier.de/rec/bibtex/#1.xml}
  {dblp:#1}}
\def\mn@eprint@#1:#2:#3:#4\@nil{\def\@tempa {#1}\def\@tempb {#2}\def\@tempc
  {#3}\ifx \@tempc \@empty \let \@tempc \@tempb \let \@tempb \@tempa \fi \ifx
  \@tempb \@empty \def\@tempb {arXiv}\fi \@ifundefined
  {mn@eprint@\@tempb}{\@tempb:\@tempc}{\expandafter \expandafter \csname
  mn@eprint@\@tempb\endcsname \expandafter{\@tempc}}}

\bibitem[\protect\citeauthoryear{{Agostino} \& {Salim}}{{Agostino} \&
  {Salim}}{2019}]{Agostino19}
{Agostino} C.~J.,  {Salim} S.,  2019, \mn@doi [\apj]
  {10.3847/1538-4357/ab1094}, \href
  {https://ui.adsabs.harvard.edu/abs/2019ApJ...876...12A} {876, 12}

\bibitem[\protect\citeauthoryear{{Aird} et~al.,}{{Aird} et~al.}{2012}]{Aird12}
{Aird} J.,  et~al., 2012, \mn@doi [\apj] {10.1088/0004-637X/746/1/90}, \href
  {https://ui.adsabs.harvard.edu/\#abs/2012ApJ...746...90A} {746, 90}

\bibitem[\protect\citeauthoryear{{Aird} et~al.,}{{Aird} et~al.}{2013}]{Aird13}
{Aird} J.,  et~al., 2013, \mn@doi [\apj] {10.1088/0004-637X/775/1/41}, \href
  {https://ui.adsabs.harvard.edu/abs/2013ApJ...775...41A} {775, 41}

\bibitem[\protect\citeauthoryear{{Aird}, {Coil}, {Georgakakis}, {Nandra},
  {Barro}  \& {P{\'e}rez-Gonz{\'a}lez}}{{Aird} et~al.}{2015}]{Aird15}
{Aird} J.,  {Coil} A.~L.,  {Georgakakis} A.,  {Nandra} K.,  {Barro} G.,
  {P{\'e}rez-Gonz{\'a}lez} P.~G.,  2015, \mn@doi [\mnras]
  {10.1093/mnras/stv1062}, \href
  {https://ui.adsabs.harvard.edu/abs/2015MNRAS.451.1892A} {451, 1892}

\bibitem[\protect\citeauthoryear{{Aird}, {Coil}  \& {Georgakakis}}{{Aird}
  et~al.}{2017}]{Aird17}
{Aird} J.,  {Coil} A.~L.,   {Georgakakis} A.,  2017, \mn@doi [\mnras]
  {10.1093/mnras/stw2932}, \href
  {https://ui.adsabs.harvard.edu/abs/2017MNRAS.465.3390A} {465, 3390}

\bibitem[\protect\citeauthoryear{{Aird}, {Coil}  \& {Georgakakis}}{{Aird}
  et~al.}{2018}]{Aird18}
{Aird} J.,  {Coil} A.~L.,   {Georgakakis} A.,  2018, \mn@doi [\mnras]
  {10.1093/mnras/stx2700}, \href
  {https://ui.adsabs.harvard.edu/\#abs/2018MNRAS.474.1225A} {474, 1225}

\bibitem[\protect\citeauthoryear{{Alexander} \& {Hickox}}{{Alexander} \&
  {Hickox}}{2012}]{AlexanderHickox12}
{Alexander} D.~M.,  {Hickox} R.~C.,  2012, \mn@doi [\nar]
  {10.1016/j.newar.2011.11.003}, \href
  {https://ui.adsabs.harvard.edu/abs/2012NewAR..56...93A} {56, 93}

\bibitem[\protect\citeauthoryear{{Astropy Collaboration} et~al.,}{{Astropy
  Collaboration} et~al.}{2013}]{astropy:2013}
{Astropy Collaboration} et~al., 2013, \mn@doi [\aap]
  {10.1051/0004-6361/201322068}, \href
  {https://ui.adsabs.harvard.edu/abs/2013A&amp;A...558A..33A} {558, A33}

\bibitem[\protect\citeauthoryear{{Astropy Collaboration} et~al.,}{{Astropy
  Collaboration} et~al.}{2018}]{astropy:2018}
{Astropy Collaboration} et~al., 2018, \mn@doi [\aj] {10.3847/1538-3881/aabc4f},
  \href {https://ui.adsabs.harvard.edu/abs/2018AJ....156..123A} {156, 123}

\bibitem[\protect\citeauthoryear{{Baldassare}, {Reines}, {Gallo}  \&
  {Greene}}{{Baldassare} et~al.}{2017}]{Baldassare17}
{Baldassare} V.~F.,  {Reines} A.~E.,  {Gallo} E.,   {Greene} J.~E.,  2017,
  \mn@doi [\apj] {10.3847/1538-4357/836/1/20}, \href
  {https://ui.adsabs.harvard.edu/abs/2017ApJ...836...20B} {836, 20}

\bibitem[\protect\citeauthoryear{{Baldry}, {Glazebrook}, {Brinkmann},
  {Ivezi{\'c}}, {Lupton}, {Nichol}  \& {Szalay}}{{Baldry}
  et~al.}{2004}]{Baldry04}
{Baldry} I.~K.,  {Glazebrook} K.,  {Brinkmann} J.,  {Ivezi{\'c}} {\v{Z}}.,
  {Lupton} R.~H.,  {Nichol} R.~C.,   {Szalay} A.~S.,  2004, \mn@doi [\apj]
  {10.1086/380092}, \href
  {https://ui.adsabs.harvard.edu/abs/2004ApJ...600..681B} {600, 681}

\bibitem[\protect\citeauthoryear{{Baldwin}, {Phillips}  \&
  {Terlevich}}{{Baldwin} et~al.}{1981}]{BaldwinPhillipsTerlevich81}
{Baldwin} J.~A.,  {Phillips} M.~M.,   {Terlevich} R.,  1981, \mn@doi [\pasp]
  {10.1086/130766}, \href {http://adsabs.harvard.edu/abs/1981PASP...93....5B}
  {93, 5}

\bibitem[\protect\citeauthoryear{{Bauer} et~al.,}{{Bauer}
  et~al.}{2013}]{Bauer13}
{Bauer} A.~E.,  et~al., 2013, \mn@doi [\mnras] {10.1093/mnras/stt1011}, \href
  {https://ui.adsabs.harvard.edu/abs/2013MNRAS.434..209B} {434, 209}

\bibitem[\protect\citeauthoryear{{Behroozi}, {Wechsler}, {Hearin}  \&
  {Conroy}}{{Behroozi} et~al.}{2019}]{Behroozi19}
{Behroozi} P.,  {Wechsler} R.~H.,  {Hearin} A.~P.,   {Conroy} C.,  2019,
  \mn@doi [\mnras] {10.1093/mnras/stz1182}, \href
  {https://ui.adsabs.harvard.edu/abs/2019MNRAS.488.3143B} {488, 3143}

\bibitem[\protect\citeauthoryear{{Bellovary}, {Brooks}, {Volonteri},
  {Governato}, {Quinn}  \& {Wadsley}}{{Bellovary} et~al.}{2013}]{Bellovary13}
{Bellovary} J.,  {Brooks} A.,  {Volonteri} M.,  {Governato} F.,  {Quinn} T.,
  {Wadsley} J.,  2013, \mn@doi [\apj] {10.1088/0004-637X/779/2/136}, \href
  {https://ui.adsabs.harvard.edu/abs/2013ApJ...779..136B} {779, 136}

\bibitem[\protect\citeauthoryear{{Birchall}, {Watson}  \& {Aird}}{{Birchall}
  et~al.}{2020}]{Birchall20}
{Birchall} K.~L.,  {Watson} M.~G.,   {Aird} J.,  2020, \mn@doi [\mnras]
  {10.1093/mnras/staa040}, \href
  {https://ui.adsabs.harvard.edu/abs/2020MNRAS.492.2268B} {492, 2268}

\bibitem[\protect\citeauthoryear{{Blanton} \& {Moustakas}}{{Blanton} \&
  {Moustakas}}{2009}]{BlantonMoustakas09}
{Blanton} M.~R.,  {Moustakas} J.,  2009, \mn@doi [\araa]
  {10.1146/annurev-astro-082708-101734}, \href
  {https://ui.adsabs.harvard.edu/abs/2009ARA&amp;A..47..159B} {47, 159}

\bibitem[\protect\citeauthoryear{{Bongiorno} et~al.,}{{Bongiorno}
  et~al.}{2012}]{Bongiorno12}
{Bongiorno} A.,  et~al., 2012, \mn@doi [\mnras]
  {10.1111/j.1365-2966.2012.22089.x}, \href
  {https://ui.adsabs.harvard.edu/abs/2012MNRAS.427.3103B} {427, 3103}

\bibitem[\protect\citeauthoryear{{Bongiorno} et~al.,}{{Bongiorno}
  et~al.}{2016}]{Bongiorno16}
{Bongiorno} A.,  et~al., 2016, \mn@doi [\aap] {10.1051/0004-6361/201527436},
  \href {https://ui.adsabs.harvard.edu/abs/2016A&amp;A...588A..78B} {588, A78}

\bibitem[\protect\citeauthoryear{{Brandt} \& {Alexander}}{{Brandt} \&
  {Alexander}}{2015}]{BrandtAlexander15}
{Brandt} W.~N.,  {Alexander} D.~M.,  2015, \mn@doi [Astronomy and Astrophysics
  Review] {10.1007/s00159-014-0081-z}, \href
  {https://ui.adsabs.harvard.edu/#abs/2015A&amp;ARv..23....1B} {23}

\bibitem[\protect\citeauthoryear{{Cann}, {Satyapal}, {Abel}, {Blecha},
  {Mushotzky}, {Reynolds}  \& {Secrest}}{{Cann} et~al.}{2019}]{Cann19}
{Cann} J.~M.,  {Satyapal} S.,  {Abel} N.~P.,  {Blecha} L.,  {Mushotzky} R.~F.,
  {Reynolds} C.~S.,   {Secrest} N.~J.,  2019, \mn@doi [\apj]
  {10.3847/2041-8213/aaf88d}, \href
  {https://ui.adsabs.harvard.edu/\#abs/2019ApJ...870L...2C} {870, L2}

\bibitem[\protect\citeauthoryear{{Carrera} et~al.,}{{Carrera}
  et~al.}{2007}]{Carrera07}
{Carrera} F.~J.,  et~al., 2007, \mn@doi [\aap] {10.1051/0004-6361:20066271},
  \href {https://ui.adsabs.harvard.edu/abs/2007A&amp;A...469...27C} {469, 27}

\bibitem[\protect\citeauthoryear{{Charlot} \& {Fall}}{{Charlot} \&
  {Fall}}{2000}]{CharlotFall00}
{Charlot} S.,  {Fall} S.~M.,  2000, \mn@doi [\apj] {10.1086/309250}, \href
  {https://ui.adsabs.harvard.edu/abs/2000ApJ...539..718C} {539, 718}

\bibitem[\protect\citeauthoryear{{Coil} et~al.,}{{Coil} et~al.}{2011}]{Coil11}
{Coil} A.~L.,  et~al., 2011, \mn@doi [\apj] {10.1088/0004-637X/741/1/8}, \href
  {https://ui.adsabs.harvard.edu/abs/2011ApJ...741....8C} {741, 8}

\bibitem[\protect\citeauthoryear{{Cool} et~al.,}{{Cool} et~al.}{2013}]{Cool13}
{Cool} R.~J.,  et~al., 2013, \mn@doi [\apj] {10.1088/0004-637X/767/2/118},
  \href {https://ui.adsabs.harvard.edu/abs/2013ApJ...767..118C} {767, 118}

\bibitem[\protect\citeauthoryear{{Delvecchio} et~al.,}{{Delvecchio}
  et~al.}{2014}]{Delvecchio14}
{Delvecchio} I.,  et~al., 2014, \mn@doi [\mnras] {10.1093/mnras/stu130}, \href
  {https://ui.adsabs.harvard.edu/abs/2014MNRAS.439.2736D} {439, 2736}

\bibitem[\protect\citeauthoryear{{Delvecchio} et~al.,}{{Delvecchio}
  et~al.}{2018}]{Delvecchio18}
{Delvecchio} I.,  et~al., 2018, \mn@doi [\mnras] {10.1093/mnras/sty2600}, \href
  {https://ui.adsabs.harvard.edu/abs/2018MNRAS.481.4971D} {481, 4971}

\bibitem[\protect\citeauthoryear{{Delvecchio} et~al.,}{{Delvecchio}
  et~al.}{2020}]{Delvecchio20}
{Delvecchio} I.,  et~al., 2020, \mn@doi [\apj] {10.3847/1538-4357/ab789c},
  \href {https://ui.adsabs.harvard.edu/abs/2020ApJ...892...17D} {892, 17}

\bibitem[\protect\citeauthoryear{{Di Matteo}, {Springel}  \& {Hernquist}}{{Di
  Matteo} et~al.}{2005}]{DiMatteo05}
{Di Matteo} T.,  {Springel} V.,   {Hernquist} L.,  2005, \mn@doi [\nat]
  {10.1038/nature03335}, \href
  {https://ui.adsabs.harvard.edu/abs/2005Natur.433..604D} {433, 604}

\bibitem[\protect\citeauthoryear{{Fabian}}{{Fabian}}{2012}]{Fabian12}
{Fabian} A.~C.,  2012, \mn@doi [\araa] {10.1146/annurev-astro-081811-125521},
  \href {https://ui.adsabs.harvard.edu/abs/2012ARA&amp;A..50..455F} {50, 455}

\bibitem[\protect\citeauthoryear{{Ferrarese} \& {Merritt}}{{Ferrarese} \&
  {Merritt}}{2000}]{FerrareseMerritt00}
{Ferrarese} L.,  {Merritt} D.,  2000, \mn@doi [\apjl] {10.1086/312838}, \href
  {https://ui.adsabs.harvard.edu/abs/2000ApJ...539L...9F} {539, L9}

\bibitem[\protect\citeauthoryear{{Gebhardt} et~al.,}{{Gebhardt}
  et~al.}{2000}]{Gebhardt00}
{Gebhardt} K.,  et~al., 2000, \mn@doi [\apjl] {10.1086/312840}, \href
  {https://ui.adsabs.harvard.edu/abs/2000ApJ...539L..13G} {539, L13}

\bibitem[\protect\citeauthoryear{{Gehrels}}{{Gehrels}}{1986}]{Gehrels86}
{Gehrels} N.,  1986, \mn@doi [\apj] {10.1086/164079}, \href
  {https://ui.adsabs.harvard.edu/\#abs/1986ApJ...303..336G} {303, 336}

\bibitem[\protect\citeauthoryear{{Georgakakis} \& {Nandra}}{{Georgakakis} \&
  {Nandra}}{2011}]{GeorgakakisNandra11}
{Georgakakis} A.,  {Nandra} K.,  2011, \mn@doi [\mnras]
  {10.1111/j.1365-2966.2011.18387.x}, \href
  {https://ui.adsabs.harvard.edu/abs/2011MNRAS.414..992G} {414, 992}

\bibitem[\protect\citeauthoryear{{Georgakakis} et~al.,}{{Georgakakis}
  et~al.}{2011}]{Georgakakis11}
{Georgakakis} A.,  et~al., 2011, \mn@doi [\mnras]
  {10.1111/j.1365-2966.2011.19650.x}, \href
  {https://ui.adsabs.harvard.edu/abs/2011MNRAS.418.2590G} {418, 2590}

\bibitem[\protect\citeauthoryear{{Georgakakis}, {Aird}, {Schulze}, {Dwelly},
  {Salvato}, {Nandra}, {Merloni}  \& {Schneider}}{{Georgakakis}
  et~al.}{2017}]{Georgakakis17}
{Georgakakis} A.,  {Aird} J.,  {Schulze} A.,  {Dwelly} T.,  {Salvato} M.,
  {Nandra} K.,  {Merloni} A.,   {Schneider} D.~P.,  2017, \mn@doi [\mnras]
  {10.1093/mnras/stx1602}, \href
  {https://ui.adsabs.harvard.edu/abs/2017MNRAS.471.1976G} {471, 1976}

\bibitem[\protect\citeauthoryear{{Greene}, {Setton}, {Bezanson}, {Suess},
  {Kriek}, {Spilker}, {Goulding}  \& {Feldmann}}{{Greene}
  et~al.}{2020}]{Greene20}
{Greene} J.~E.,  {Setton} D.,  {Bezanson} R.,  {Suess} K.~A.,  {Kriek} M.,
  {Spilker} J.~S.,  {Goulding} A.~D.,   {Feldmann} R.,  2020, \mn@doi [\apjl]
  {10.3847/2041-8213/aba534}, \href
  {https://ui.adsabs.harvard.edu/abs/2020ApJ...899L...9G} {899, L9}

\bibitem[\protect\citeauthoryear{{Haggard}, {Green}, {Anderson}, {Constantin},
  {Aldcroft}, {Kim}  \& {Barkhouse}}{{Haggard} et~al.}{2010}]{Haggard10}
{Haggard} D.,  {Green} P.~J.,  {Anderson} S.~F.,  {Constantin} A.,  {Aldcroft}
  T.~L.,  {Kim} D.-W.,   {Barkhouse} W.~A.,  2010, \mn@doi [\apj]
  {10.1088/0004-637X/723/2/1447}, \href
  {https://ui.adsabs.harvard.edu/abs/2010ApJ...723.1447H} {723, 1447}

\bibitem[\protect\citeauthoryear{{Hern{\'a}n-Caballero}
  et~al.,}{{Hern{\'a}n-Caballero} et~al.}{2014}]{HernanCaballero14}
{Hern{\'a}n-Caballero} A.,  et~al., 2014, \mn@doi [\mnras]
  {10.1093/mnras/stu1413}, \href
  {https://ui.adsabs.harvard.edu/abs/2014MNRAS.443.3538H} {443, 3538}

\bibitem[\protect\citeauthoryear{{Hickox}, {Mullaney}, {Alexander}, {Chen},
  {Civano}, {Goulding}  \& {Hainline}}{{Hickox} et~al.}{2014}]{Hickox14}
{Hickox} R.~C.,  {Mullaney} J.~R.,  {Alexander} D.~M.,  {Chen} C.-T.~J.,
  {Civano} F.~M.,  {Goulding} A.~D.,   {Hainline} K.~N.,  2014, \mn@doi [\apj]
  {10.1088/0004-637X/782/1/9}, \href
  {https://ui.adsabs.harvard.edu/abs/2014ApJ...782....9H} {782, 9}

\bibitem[\protect\citeauthoryear{{Jahnke} et~al.,}{{Jahnke}
  et~al.}{2009}]{Jahnke09}
{Jahnke} K.,  et~al., 2009, \mn@doi [\apjl] {10.1088/0004-637X/706/2/L215},
  \href {https://ui.adsabs.harvard.edu/abs/2009ApJ...706L.215J} {706, L215}

\bibitem[\protect\citeauthoryear{{Jones}, {Hickox}, {Mutch}, {Croton}, {Ptak}
  \& {DiPompeo}}{{Jones} et~al.}{2017}]{Jones17}
{Jones} M.~L.,  {Hickox} R.~C.,  {Mutch} S.~J.,  {Croton} D.~J.,  {Ptak} A.~F.,
    {DiPompeo} M.~A.,  2017, \mn@doi [\apj] {10.3847/1538-4357/aa7632}, \href
  {https://ui.adsabs.harvard.edu/abs/2017ApJ...843..125J} {843, 125}

\bibitem[\protect\citeauthoryear{{Kauffmann} et~al.,}{{Kauffmann}
  et~al.}{2003a}]{Kauffmann03b}
{Kauffmann} G.,  et~al., 2003a, \mn@doi [\mnras]
  {10.1111/j.1365-2966.2003.07154.x}, \href
  {http://adsabs.harvard.edu/abs/2003MNRAS.346.1055K} {346, 1055}

\bibitem[\protect\citeauthoryear{{Kauffmann} et~al.,}{{Kauffmann}
  et~al.}{2003b}]{Kauffmann03}
{Kauffmann} G.,  et~al., 2003b, \mn@doi [\mnras]
  {10.1111/j.1365-2966.2003.07154.x}, \href
  {https://ui.adsabs.harvard.edu/abs/2003MNRAS.346.1055K} {346, 1055}

\bibitem[\protect\citeauthoryear{{Kewley}, {Dopita}, {Sutherland}, {Heisler}
  \& {Trevena}}{{Kewley} et~al.}{2001}]{Kewley01}
{Kewley} L.~J.,  {Dopita} M.~A.,  {Sutherland} R.~S.,  {Heisler} C.~A.,
  {Trevena} J.,  2001, \mn@doi [\apj] {10.1086/321545}, \href
  {http://adsabs.harvard.edu/abs/2001ApJ...556..121K} {556, 121}

\bibitem[\protect\citeauthoryear{{Kormendy} \& {Ho}}{{Kormendy} \&
  {Ho}}{2013}]{KormendyHo13}
{Kormendy} J.,  {Ho} L.~C.,  2013, \mn@doi [Annual Review of Astronomy and
  Astrophysics] {10.1146/annurev-astro-082708-101811}, \href
  {https://ui.adsabs.harvard.edu/#abs/2013ARA&amp;A..51..511K} {51, 511}

\bibitem[\protect\citeauthoryear{{Lehmer} et~al.,}{{Lehmer}
  et~al.}{2016}]{Lehmer16}
{Lehmer} B.~D.,  et~al., 2016, \mn@doi [\apj] {10.3847/0004-637X/825/1/7},
  \href {http://adsabs.harvard.edu/abs/2016ApJ...825....7L} {825, 7}

\bibitem[\protect\citeauthoryear{{Lemons}, {Reines}, {Plotkin}, {Gallo}  \&
  {Greene}}{{Lemons} et~al.}{2015}]{Lemons15}
{Lemons} S.~M.,  {Reines} A.~E.,  {Plotkin} R.~M.,  {Gallo} E.,   {Greene}
  J.~E.,  2015, \mn@doi [\apj] {10.1088/0004-637X/805/1/12}, \href
  {https://ui.adsabs.harvard.edu/#abs/2015ApJ...805...12L} {805}

\bibitem[\protect\citeauthoryear{{Lusso} \& {Risaliti}}{{Lusso} \&
  {Risaliti}}{2016}]{LussoRisaliti16}
{Lusso} E.,  {Risaliti} G.,  2016, \mn@doi [\apj]
  {10.3847/0004-637X/819/2/154}, \href
  {http://adsabs.harvard.edu/abs/2016ApJ...819..154L} {819, 154}

\bibitem[\protect\citeauthoryear{{Madau} \& {Dickinson}}{{Madau} \&
  {Dickinson}}{2014}]{MadauDickinson14}
{Madau} P.,  {Dickinson} M.,  2014, \mn@doi [\araa]
  {10.1146/annurev-astro-081811-125615}, \href
  {https://ui.adsabs.harvard.edu/abs/2014ARA&amp;A..52..415M} {52, 415}

\bibitem[\protect\citeauthoryear{{Magorrian} et~al.,}{{Magorrian}
  et~al.}{1998}]{Magorrian98}
{Magorrian} J.,  et~al., 1998, \mn@doi [\aj] {10.1086/300353}, \href
  {https://ui.adsabs.harvard.edu/abs/1998AJ....115.2285M} {115, 2285}

\bibitem[\protect\citeauthoryear{{Martin} et~al.,}{{Martin}
  et~al.}{2005}]{Martin05}
{Martin} D.~C.,  et~al., 2005, \mn@doi [\apjl] {10.1086/426387}, \href
  {https://ui.adsabs.harvard.edu/abs/2005ApJ...619L...1M} {619, L1}

\bibitem[\protect\citeauthoryear{{Martin} et~al.,}{{Martin}
  et~al.}{2007}]{Martin07}
{Martin} D.~C.,  et~al., 2007, \mn@doi [\apjs] {10.1086/516639}, \href
  {https://ui.adsabs.harvard.edu/abs/2007ApJS..173..342M} {173, 342}

\bibitem[\protect\citeauthoryear{{Mendez} et~al.,}{{Mendez}
  et~al.}{2013}]{Mendez13}
{Mendez} A.~J.,  et~al., 2013, \mn@doi [\apj] {10.1088/0004-637X/770/1/40},
  \href {https://ui.adsabs.harvard.edu/abs/2013ApJ...770...40M} {770, 40}

\bibitem[\protect\citeauthoryear{Mezcua, Civano, Marchesi, Suh, Fabbiano  \&
  Volonteri}{Mezcua et~al.}{2018}]{Mezcua18}
Mezcua M.,  Civano F.,  Marchesi S.,  Suh H.,  Fabbiano G.,   Volonteri M.,
  2018, \mn@doi [Monthly Notices of the Royal Astronomical Society]
  {10.1093/mnras/sty1163}, 478, 2576

\bibitem[\protect\citeauthoryear{{Mineo}, {Gilfanov}  \& {Sunyaev}}{{Mineo}
  et~al.}{2012a}]{Mineo12a}
{Mineo} S.,  {Gilfanov} M.,   {Sunyaev} R.,  2012a, \mn@doi [\mnras]
  {10.1111/j.1365-2966.2011.19862.x}, \href
  {https://ui.adsabs.harvard.edu/abs/2012MNRAS.419.2095M} {419, 2095}

\bibitem[\protect\citeauthoryear{{Mineo}, {Gilfanov}  \& {Sunyaev}}{{Mineo}
  et~al.}{2012b}]{Mineo12b}
{Mineo} S.,  {Gilfanov} M.,   {Sunyaev} R.,  2012b, \mn@doi [\mnras]
  {10.1111/j.1365-2966.2012.21831.x}, \href
  {http://adsabs.harvard.edu/abs/2012MNRAS.426.1870M} {426, 1870}

\bibitem[\protect\citeauthoryear{{Moustakas} et~al.,}{{Moustakas}
  et~al.}{2013}]{Moustakas13}
{Moustakas} J.,  et~al., 2013, \mn@doi [\apj] {10.1088/0004-637X/767/1/50},
  \href {https://ui.adsabs.harvard.edu/abs/2013ApJ...767...50M} {767, 50}

\bibitem[\protect\citeauthoryear{{Mullaney} et~al.,}{{Mullaney}
  et~al.}{2012}]{Mullaney12}
{Mullaney} J.~R.,  et~al., 2012, \mn@doi [\apjl] {10.1088/2041-8205/753/2/L30},
  \href {https://ui.adsabs.harvard.edu/abs/2012ApJ...753L..30M} {753, L30}

\bibitem[\protect\citeauthoryear{{Muzzin} et~al.,}{{Muzzin}
  et~al.}{2013}]{Muzzin13}
{Muzzin} A.,  et~al., 2013, \mn@doi [\apj] {10.1088/0004-637X/777/1/18}, \href
  {https://ui.adsabs.harvard.edu/abs/2013ApJ...777...18M} {777, 18}

\bibitem[\protect\citeauthoryear{{Paggi}, {Fabbiano}, {Civano}, {Pellegrini},
  {Elvis}  \& {Kim}}{{Paggi} et~al.}{2016}]{Paggi16}
{Paggi} A.,  {Fabbiano} G.,  {Civano} F.,  {Pellegrini} S.,  {Elvis} M.,
  {Kim} D.-W.,  2016, \mn@doi [\apj] {10.3847/0004-637X/823/2/112}, \href
  {https://ui.adsabs.harvard.edu/#abs/2016ApJ...823..112P} {823}

\bibitem[\protect\citeauthoryear{{Pardo} et~al.,}{{Pardo}
  et~al.}{2016}]{Pardo16}
{Pardo} K.,  et~al., 2016, \mn@doi [\apj] {10.3847/0004-637X/831/2/203}, \href
  {https://ui.adsabs.harvard.edu/#abs/2016ApJ...831..203P} {831}

\bibitem[\protect\citeauthoryear{{Pineau} et~al.,}{{Pineau}
  et~al.}{2017}]{Pineau17}
{Pineau} F.~X.,  et~al., 2017, \mn@doi [\aap] {10.1051/0004-6361/201629219},
  \href {https://ui.adsabs.harvard.edu/abs/2017A&amp;A...597A..89P} {597, A89}

\bibitem[\protect\citeauthoryear{{Popping}, {Caputi}, {Somerville}  \&
  {Trager}}{{Popping} et~al.}{2012}]{Popping12}
{Popping} G.,  {Caputi} K.~I.,  {Somerville} R.~S.,   {Trager} S.~C.,  2012,
  \mn@doi [\mnras] {10.1111/j.1365-2966.2012.21702.x}, \href
  {https://ui.adsabs.harvard.edu/abs/2012MNRAS.425.2386P} {425, 2386}

\bibitem[\protect\citeauthoryear{{Rosen} et~al.,}{{Rosen}
  et~al.}{2016}]{Rosen16}
{Rosen} S.~R.,  et~al., 2016, \mn@doi [\aap] {10.1051/0004-6361/201526416},
  \href {http://adsabs.harvard.edu/abs/2016A%26A...590A...1R} {590, A1}

\bibitem[\protect\citeauthoryear{{Salim} et~al.,}{{Salim}
  et~al.}{2007}]{Salim07}
{Salim} S.,  et~al., 2007, \mn@doi [\apjs] {10.1086/519218}, \href
  {http://adsabs.harvard.edu/abs/2007ApJS..173..267S} {173, 267}

\bibitem[\protect\citeauthoryear{{Salim} et~al.,}{{Salim}
  et~al.}{2016}]{Salim16}
{Salim} S.,  et~al., 2016, \mn@doi [\apjs] {10.3847/0067-0049/227/1/2}, \href
  {https://ui.adsabs.harvard.edu/abs/2016ApJS..227....2S} {227, 2}

\bibitem[\protect\citeauthoryear{{Schawinski} et~al.,}{{Schawinski}
  et~al.}{2014}]{Schawinski14}
{Schawinski} K.,  et~al., 2014, \mn@doi [\mnras] {10.1093/mnras/stu327}, \href
  {https://ui.adsabs.harvard.edu/abs/2014MNRAS.440..889S} {440, 889}

\bibitem[\protect\citeauthoryear{{Vanden Berk} et~al.,}{{Vanden Berk}
  et~al.}{2001}]{VandenBerk01}
{Vanden Berk} D.~E.,  et~al., 2001, \mn@doi [\aj] {10.1086/321167}, \href
  {http://adsabs.harvard.edu/abs/2001AJ....122..549V} {122, 549}

\bibitem[\protect\citeauthoryear{{Vito} et~al.,}{{Vito} et~al.}{2014}]{Vito14}
{Vito} F.,  et~al., 2014, \mn@doi [\mnras] {10.1093/mnras/stu637}, \href
  {https://ui.adsabs.harvard.edu/abs/2014MNRAS.441.1059V} {441, 1059}

\bibitem[\protect\citeauthoryear{{Wang} et~al.,}{{Wang} et~al.}{2017}]{Wang17}
{Wang} T.,  et~al., 2017, \mn@doi [\aap] {10.1051/0004-6361/201526645}, \href
  {https://ui.adsabs.harvard.edu/abs/2017A&amp;A...601A..63W} {601, A63}

\bibitem[\protect\citeauthoryear{{Weigel}, {Schawinski}, {Caplar}, {Wong},
  {Treister}  \& {Trakhtenbrot}}{{Weigel} et~al.}{2017}]{Weigel17}
{Weigel} A.~K.,  {Schawinski} K.,  {Caplar} N.,  {Wong} O.~I.,  {Treister} E.,
   {Trakhtenbrot} B.,  2017, \mn@doi [\apj] {10.3847/1538-4357/aa803b}, \href
  {https://ui.adsabs.harvard.edu/abs/2017ApJ...845..134W} {845, 134}

\bibitem[\protect\citeauthoryear{{Williams} \& {R{\"o}ttgering}}{{Williams} \&
  {R{\"o}ttgering}}{2015}]{WilliamsRottgering15}
{Williams} W.~L.,  {R{\"o}ttgering} H.~J.~A.,  2015, \mn@doi [\mnras]
  {10.1093/mnras/stv692}, \href
  {https://ui.adsabs.harvard.edu/abs/2015MNRAS.450.1538W} {450, 1538}

\bibitem[\protect\citeauthoryear{{Xue} et~al.,}{{Xue} et~al.}{2010}]{Xue10}
{Xue} Y.~Q.,  et~al., 2010, \mn@doi [\apj] {10.1088/0004-637X/720/1/368}, \href
  {https://ui.adsabs.harvard.edu/abs/2010ApJ...720..368X} {720, 368}

\bibitem[\protect\citeauthoryear{{Yang} et~al.,}{{Yang} et~al.}{2017}]{Yang17}
{Yang} G.,  et~al., 2017, \mn@doi [\apj] {10.3847/1538-4357/aa7564}, \href
  {https://ui.adsabs.harvard.edu/abs/2017ApJ...842...72Y} {842, 72}

\bibitem[\protect\citeauthoryear{{Yang} et~al.,}{{Yang} et~al.}{2018}]{Yang18}
{Yang} G.,  et~al., 2018, \mn@doi [\mnras] {10.1093/mnras/stx2805}, \href
  {https://ui.adsabs.harvard.edu/abs/2018MNRAS.475.1887Y} {475, 1887}

\makeatother
\end{thebibliography}

\appendix

\onecolumn
\section{Probability Distribution Fit Coefficients}
Table A1 outlines the best-fit coefficients, and associated errors, for equation \eqref{eq:powerlaw} to create every probability distribution shown in figure \ref{fig:Prob_Dist_Comparison}. 

\label{tab:Prob_Dist_Coeffs}

\begin{longtable}{|l|cc|cc|}

\hline

\multicolumn{1}{|c|}{} & \multicolumn{2}{c|}{sBHAR}
& \multicolumn{2}{c|}{X-ray Luminosity ($\mathrm{erg/s}$)} \\

\multicolumn{1}{|c|}{} & \multicolumn{2}{c|}{($\log_{10}\ x' = -2.55$)}
& \multicolumn{2}{c|}{($\log_{10}\ x' = 42.1$)}\\

\hline

z & $\log_{10}\ A$ & $k$ & $\log_{10}\ A$ & $k$ \\

\hline

0.00 - 0.05 & $-2.44 \pm 0.11$ & $-0.57 \pm 0.13$ & $-2.73 \pm 0.13$ & $-0.43 \pm 0.12$ \\

\rowcolor{Gray}
0.05 - 0.10 & $-2.48 \pm 0.08$ & $-0.72 \pm 0.11$ & $-2.65 \pm 0.10$ & $-0.97 \pm 0.12$  \\

0.10 - 0.15 & $-2.54 \pm 0.12$ & $-1.39 \pm 0.14$ & $-2.34 \pm 0.10$ & $-1.78 \pm 0.19$ \\

\rowcolor{Gray}
0.15 - 0.20 & $-2.36 \pm 0.13$ & $-1.23 \pm 0.18$ & $-1.72 \pm 0.10$ & $-1.34 \pm 0.20$ \\
    
0.20 - 0.25 & $-2.13 \pm 0.17$ & $-1.15 \pm 0.23$ & $-1.24 \pm 0.15$ & $-1.64 \pm 0.32$ \\
\rowcolor{Gray}
0.25 - 0.35 & $-2.08 \pm 0.19$ & $-1.59 \pm 0.24$ & $-0.64 \pm 0.20$ & $-2.01 \pm 0.39$ \\

\hline

$\log_{10}$ Stellar Mass ($\mathrm{M_\odot}$) & $\log_{10}\ A$ & $k$ & $\log_{10}\ A$ & $k$ \\

\hline
\rowcolor{Gray}
8.00 - 9.00 & $-2.24 \pm 0.61$ & $-0.90 \pm 0.01$ & $-3.05 \pm 0.00$ & $-0.37 \pm 0.43$ \\

9.00 - 10.00 & $-2.61 \pm 0.22$ & $-0.90 \pm 0.26$ & $-3.24 \pm 0.43$ & $-0.74 \pm 0.25$ \\
\rowcolor{Gray}
10.00 - 10.50 & $-2.49 \pm 0.11$ & $-0.49 \pm 0.16$ & $-2.65 \pm 0.14$ & $-0.50 \pm 0.16$ \\

10.50 - 11.00 & $-2.23 \pm 0.06$ & $-0.52 \pm 0.08$ & $-2.16 \pm 0.05$ & $-0.48 \pm 0.08$ \\
\rowcolor{Gray}
11.00 - 11.50 & $-2.63 \pm 0.12$ & $-1.22 \pm 0.13$ & $-1.91 \pm 0.07$ & $-1.07 \pm 0.12$ \\

11.50 - 12.00 & $-2.39 \pm 0.25$ & $-1.11 \pm 0.26$ & $-1.31 \pm 0.14$ & $-1.12 \pm 0.27$ \\

\hline

\multicolumn{5}{|l|}{\textbf{Table A1:} Best-fit coefficients used to fit equation \eqref{eq:powerlaw} to all probability distribution configurations.}\\

\hline

\end{longtable}

\section{Other Probability Distributions}
\label{app:Prob_Dists}
Figures \ref{fig:Other_Prob_Dists_1} - \ref{fig:Other_Prob_Dists_3} show the probability distributions split by stellar mass and redshift, and the remaining trends with X-ray luminosity and sBHAR. They have the same form as figure \ref{fig:Example_Prob_Dist} and are included for the sake of transparency, to show the strength of our power law fits to the data. 

\begin{figure*}

    \centering
    \includegraphics[width=0.8\columnwidth]{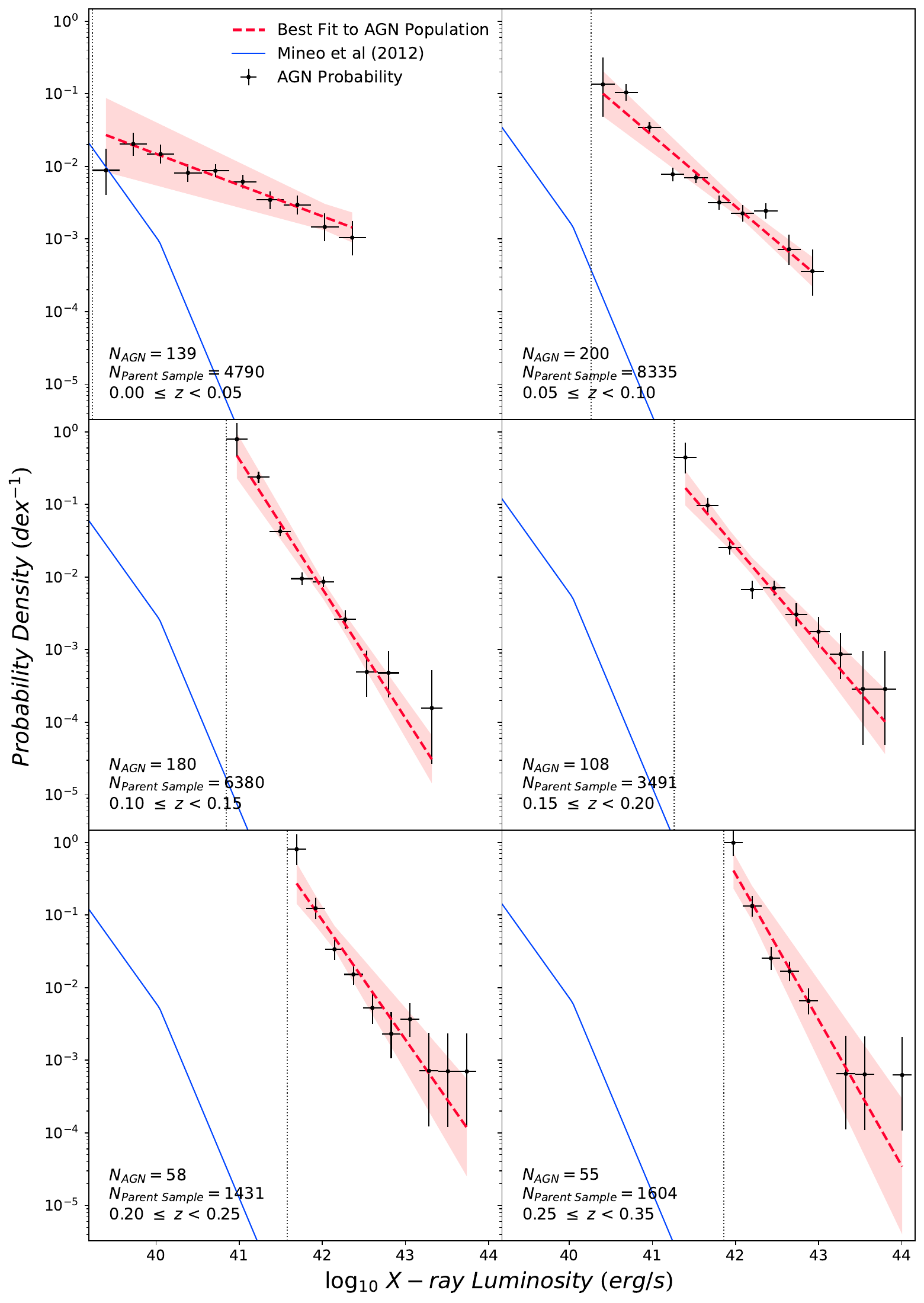}

    \caption[Additional MPA-JHU AGN probability distributions: X-ray luminosity \& redshift]{Additional probability distribution used to calculate fits and associated error regions for the other host galaxy properties. This figure shows the probability of a galaxy hosting an AGN as a function of X-ray luminosity and split into bins of redshift. As with figure \ref{fig:Example_Prob_Dist}, power laws (dashed red lines) have been fit to the data in each panel and displayed alongside their $1\sigma$ uncertainty (pale red region). These plots were constructed using the method outlined in section \ref{sec:Prob_Dists}}
    \label{fig:Other_Prob_Dists_1}
\end{figure*}

\begin{figure*}
    \ContinuedFloat
    \centering
    \includegraphics[width=0.8\columnwidth]{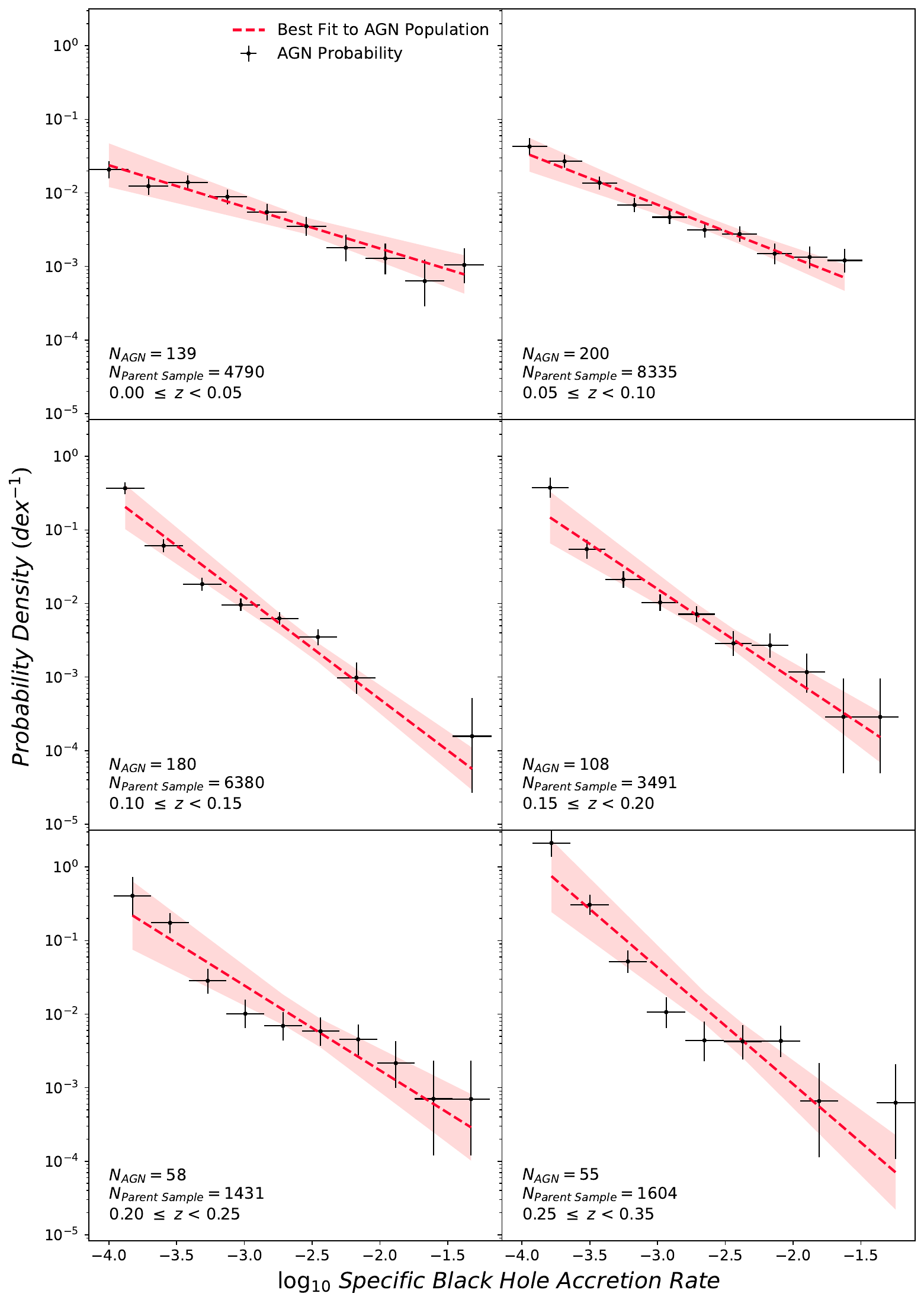}

    \caption[Additional MPA-JHU AGN probability distributions: $\mathrm{\lambda_{sBHAR}}$ \& redshift]{Additional probability distribution used to calculate fits and associated error regions for the other host galaxy properties. This figure shows the probability of a galaxy hosting an AGN as a function of $\mathrm{\lambda_{sBHAR}}$ and split into bins of redshift. As with figure \ref{fig:Example_Prob_Dist}, power laws (dashed red lines) have been fit to the data in each panel and displayed alongside their $1\sigma$ uncertainty (pale red region). These plots were constructed using the method outlined in section \ref{sec:Prob_Dists}}
    \label{fig:Other_Prob_Dists_2}
\end{figure*}

\begin{figure*}
    \ContinuedFloat
    \centering
    \includegraphics[width=0.8\columnwidth]{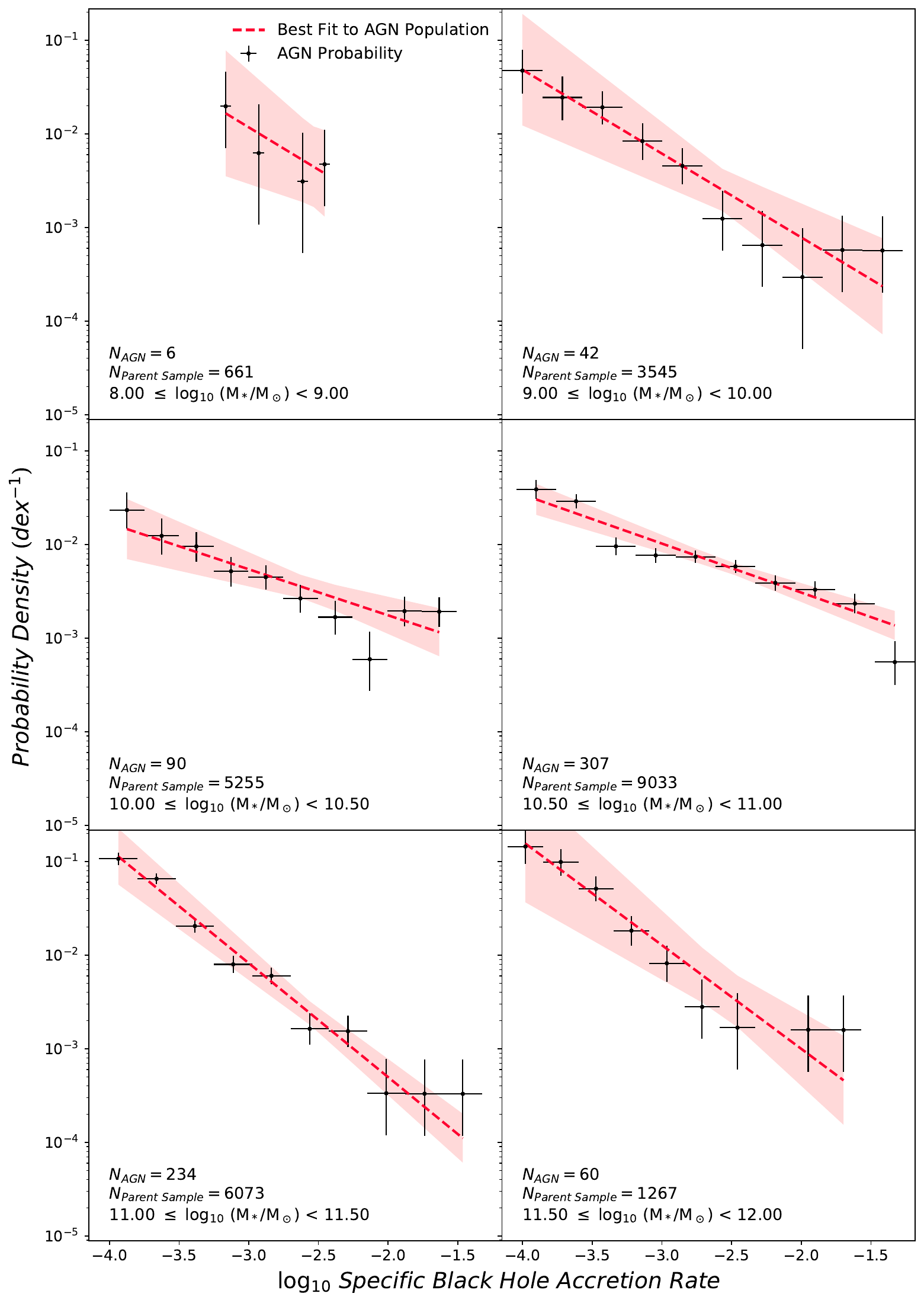}
    
    \caption[Additional MPA-JHU AGN probability distributions: $\mathrm{\lambda_{sBHAR}}$ \& stellar mass]{Additional probability distribution used to calculate fits and associated error regions for the other host galaxy properties. This figure shows the probability of a galaxy hosting an AGN as a function of $\mathrm{\lambda_{sBHAR}}$ and split into bins of stellar mass. As with figure \ref{fig:Example_Prob_Dist}, power laws (dashed red lines) have been fit to the data in each panel and displayed alongside their $1\sigma$ uncertainty (pale red region). These plots were constructed using the method outlined in section \ref{sec:Prob_Dists}}
    \label{fig:Other_Prob_Dists_3}
\end{figure*}

\bsp
\label{lastpage}
\end{document}